\documentclass[fleqn,usenatbib]{mnras}

\usepackage{newtxtext,newtxmath}

\usepackage[T1]{fontenc}
\usepackage{ae,aecompl}

\usepackage{graphicx}	
\usepackage{amsmath}	
\usepackage{amssymb}	
\usepackage{natbib}

\def\M{M$_\odot$}
\def\ergs{erg\,s$^{-1}$}
\def\kms{km\,s$^{-1}$}

\title[A TDE changes its spots]{The tidal disruption event AT2017eqx: spectroscopic evolution from hydrogen rich to poor suggests an atmosphere and outflow}

\author[M. Nicholl et al.]
{M.~Nicholl,$^{1,2}$\thanks{E-mail: mrn@roe.ac.uk} P.~K.~Blanchard,$^{3}$ E.~Berger,$^{3}$ S.~Gomez,$^{3}$ R.~Margutti,$^{4}$
\newauthor K.~D.~Alexander,$^{4,5}$ J.~Guillochon,$^{3}$ J.~Leja,$^{3,6}$ R.~Chornock,$^{7}$ B.~Snios,$^{3}$
\newauthor K.~Auchettl,$^{8,9}$ A.~G.~Bruce,$^{1}$ P.~Challis,$^{3}$ D.~J.~D'Orazio,$^{3}$ M.~R.~Drout,$^{10,11}$
\newauthor T.~Eftekhari,$^{3}$ R.~J.~Foley,$^{12}$ O.~Graur,$^{3,6,13}$ C.~D.~Kilpatrick,$^{12}$ A.~Lawrence,$^{1}$
\newauthor A.~L.~Piro,$^{10}$ C.~Rojas-Bravo,$^{12}$ N.~P.~Ross,$^{1}$ P.~Short,$^{1}$ S.~J.~Smartt,$^{14}$
\newauthor K.~W.~Smith,$^{14}$ B.~Stalder$^{15}$
\\
$^{1}$Institute for Astronomy, University of Edinburgh, Royal Observatory, Blackford Hill, EH9 3HJ, UK \\
$^{2}$Birmingham Institute for Gravitational Wave Astronomy and School of Physics and Astronomy, University of Birmingham, \\ Birmingham B15 2TT, UK \\
$^{3}$Harvard-Smithsonian Center for Astrophysics, 60 Garden Street, Cambridge,Massachusetts, 02138, USA \\
$^{4}$Center for Interdisciplinary Exploration and Research in Astrophysics (CIERA) and Department of Physics and Astronomy, \\ Northwestern University, Evanston, IL 60208, USA \\
$^{5}$Einstein Fellow \\
$^{6}$NSF Astronomy and Astrophysics Postdoctoral Fellow \\
$^{7}$Astrophysical Institute, Department of Physics and Astronomy, 251B Clippinger Lab, Ohio University, Athens, OH 45701, USA \\
$^{8}$Center for Cosmology and Astro-Particle Physics and Department of Physics, The Ohio State University, 191 West Woodruff \\ Avenue, Columbus, \\ OH 43210, USA \\
$^{9}$DARK, Niels Bohr Institute, University of Copenhagen, Lyngbyvej 2, 2100 Copenhagen, Denmark \\
$^{10}$The Observatories of the Carnegie Institution for Science, 813 Santa Barbara St., Pasadena, CA 91101, USA \\
$^{11}$Department of Astronomy and Astrophysics, University of Toronto, 50 St.~George St., Toronto, Ontario, M5S 3H4 Canada \\
$^{12}$Department of Astronomy and Astrophysics, University of California, Santa Cruz, CA 95064, USA \\
$^{13}$Department of Astrophysics, American Museum of Natural History, New York, NY 10024, USA \\
$^{14}$Astrophysics Research Centre, School of Mathematics and Physics, Queens University Belfast, Belfast BT7 1NN, UK \\
$^{15}$Institute for Astronomy, University of Hawaii, 2680 Woodlawn Drive, Honolulu, HI 96822, USA \\
}

\date{Accepted XXX. Received YYY; in original form ZZZ}

\pubyear{2019}

\begin{document}
\label{firstpage}
\pagerange{\pageref{firstpage}--\pageref{lastpage}}
\maketitle

\begin{abstract}
We present and analyse a new tidal disruption event (TDE), AT2017eqx at redshift $z=0.1089$, discovered by Pan-STARRS and ATLAS. The position of the transient is consistent with the nucleus of its host galaxy; the spectrum shows a persistent blackbody temperature $T \gtrsim 20,000$\,K with broad H\,I and He\,II emission; and it peaks at a blackbody luminosity of $L \approx 10^{44}$\,\ergs. The lines are initially centered at zero velocity, but by 100 days the H\,I lines disappear while the He\,II develops a blueshift of $\gtrsim 5,000$\,\kms. Both the early- and late-time morphologies have been seen in other TDEs, but the complete transition between them is unprecedented. The evolution can be explained by combining an extended atmosphere, undergoing slow contraction, with a wind in the polar direction becoming visible at late times. Our observations confirm that a lack of hydrogen a TDE spectrum does not indicate a stripped star, while the proposed model implies that much of the diversity in TDEs may be due to the observer viewing angle. Modelling the light curve suggests AT2017eqx resulted from the complete disruption of a solar-mass star by a black hole of $\sim 10^{6.3}$\,M$_\odot$. The host is another Balmer-strong absorption galaxy, though fainter and less centrally concentrated than most TDE hosts. Radio limits rule out a relativistic jet, while X-ray limits at 500 days are among the deepest for a TDE at this phase. 
\end{abstract}




\section{Introduction}

A tidal disruption event (TDE) occurs when an unfortunate star passes so close to a supermassive black hole (SMBH) that the tidal force of the SMBH exceeds the self-gravity of the star \citep{Hills1975}. If this takes place outside of the Schwarzschild radius, the result is a luminous flare with $L_{\rm bol}\sim 10^{41-45}$\,\ergs, powered either by accretion onto the SMBH \citep{Kochanek1994,Jiang2016,Lodato2012,Piran2015}. Observationally, these are differentiated from more common transients like supernovae by their higher blackbody temperatures ($T \sim 20,000-50,000$\,K) and coincidence with the centres of galaxies.

Although TDEs were initially expected to peak at X-ray wavelengths if they are powered by accretion \citep{Komossa2002}, TDE candidates have now been discovered in the rest-frame UV and optical by surveys such as SDSS \citep{vanVelzen2011}, Pan-STARRS \citep{Gezari2012,Chornock2014,Blanchard2017,Holoien2018}, ASASSN \citep{Holoien2014,Holoien2016,Holoien2016b}, OGLE \citep{Wyrz2017}, PTF \citep{Arcavi2014,Hung2017,Blagorodnova2017,Blagorodnova2018} and ZTF \citep{vanVelzen2018}. Up to $\sim 50\%$ of these events are in fact faint in X-rays \citep{Auchettl2017}, suggesting either that some events are not powered by direct accretion, or that X-rays can only escape along certain sight-lines \citep{Dai2018}.

TDEs have been found in galaxies with stellar masses ranging from $\sim 10^{8.5}-10^{11}$\,\M\ \citep{vanVelzen2018b,Wevers2017,Wevers2019}, corresponding to black hole masses $\sim 10^{6}-10^{8}$\,\M\ \citep{Mockler2018}, and even greater in the case of rapidly rotating SMBHs \citep{Leloudas2016}. They provide a novel way to probe the properties of otherwise dormant SMBHs and their environments, especially at the lower end of the mass spectrum.  TDEs are more common in galaxies with a high stellar mass surface density and a centrally-concentrated light profile \citep{LawSmith2017,Graur2018}, but there remains an unexplained over-representation of Balmer-strong absorption galaxies \citep{Arcavi2014,French2016,French2018}.

Studying TDEs is complicated by the wide diversity in their observed characteristics, in the optical, X-rays and radio \citep{Auchettl2017,Zauderer2011,Levan2011,vanVelzen2016,Alexander2016}. Their optical spectra can exhibit lines of H\,I, He\,II, or both \citep{Arcavi2014}, but usually retain their spectroscopic signatures over time. 
A few instead show broad absorption lines \citep{Chornock2014,Leloudas2016}. More recently, metal lines have been detected in some TDEs \citep{Blagorodnova2018,Brown2018,Leloudas2019,Wevers2019b}, while yet others show spectra dominated by pre-existing broad-line and narrow-line regions from an active galactic nucleus \citep{Blanchard2017,Kankare2017}. Determining the nature of line formation in TDEs is key to understanding the physical processes in these events \citep[e.g.][]{Guillochon2014,Roth2018}.

In this paper, we present a new TDE, AT2017eqx, which we observed to undergo a radical evolution in its spectroscopic properties over time. The spectrum initially showed prominent Balmer and He\,II emission lines centred at close to zero velocity. Later spectra showed no evidence for hydrogen emission, while the He\,II feature became blueshifted by $\gtrsim 5000$\,\kms. Both of these morphologies have been observed before in other TDEs, but to our knowledge such a transition within the same event has not. Understanding this evolution can shed new light on the geometry and line-formation in these events.

Our study is organised as follows. We describe the discovery of AT2017eqx in section \ref{sec:class} and detail our multiwavelength follow-up observations in section \ref{sec:obs}. We analyze the light curve in section \ref{sec:phot} to derive bolometric properties and infer physical parameters. We present and interpret the surprising spectroscopic evolution in section \ref{sec:spec}. In section \ref{sec:host}, we study the host galaxy in the context of other TDE hosts, before concluding in section \ref{sec:conc}.

\section{Discovery and classification}\label{sec:class}

AT2017eqx (survey name, PS17dhz) was discovered by the PanSTARRS Survey for Transients \citep{Chambers2016} on 2017-06-07 UT, at right ascension 22h\,26m\,48.370s, declination 17$^\circ$\,08'\,52.40''. The source was coincident with the centre of a galaxy catalogued in the Sloan Digital Sky Survey as SDSS J222648.38+170852.2, with an apparent magnitude of $g=20.99$\,mag in Data Release 14 \citep{Abolfathi2018}. We determined the offset of AT2017eqx from the nucleus of this galaxy using a deep $g$-band image of the transient obtained with LDSS3 on 2017-07-22 and an archival pre-disruption image from PanSTARRS DR1 \citep{Flewelling2016}. We geometrically aligned these images, with 30 common stars for reference, using the \textsc{geomap} and \textsc{geotran} tasks in \textsc{pyraf}. After transforming to a common coordinate grid and measuring the centroids of the transient and host galaxy, we find a relative offset of $0.06''\pm0.08''$. Thus the origin of the source is fully consistent with the nucleus of the galaxy.

\begin{figure}
\centering
	\includegraphics[width=\columnwidth]{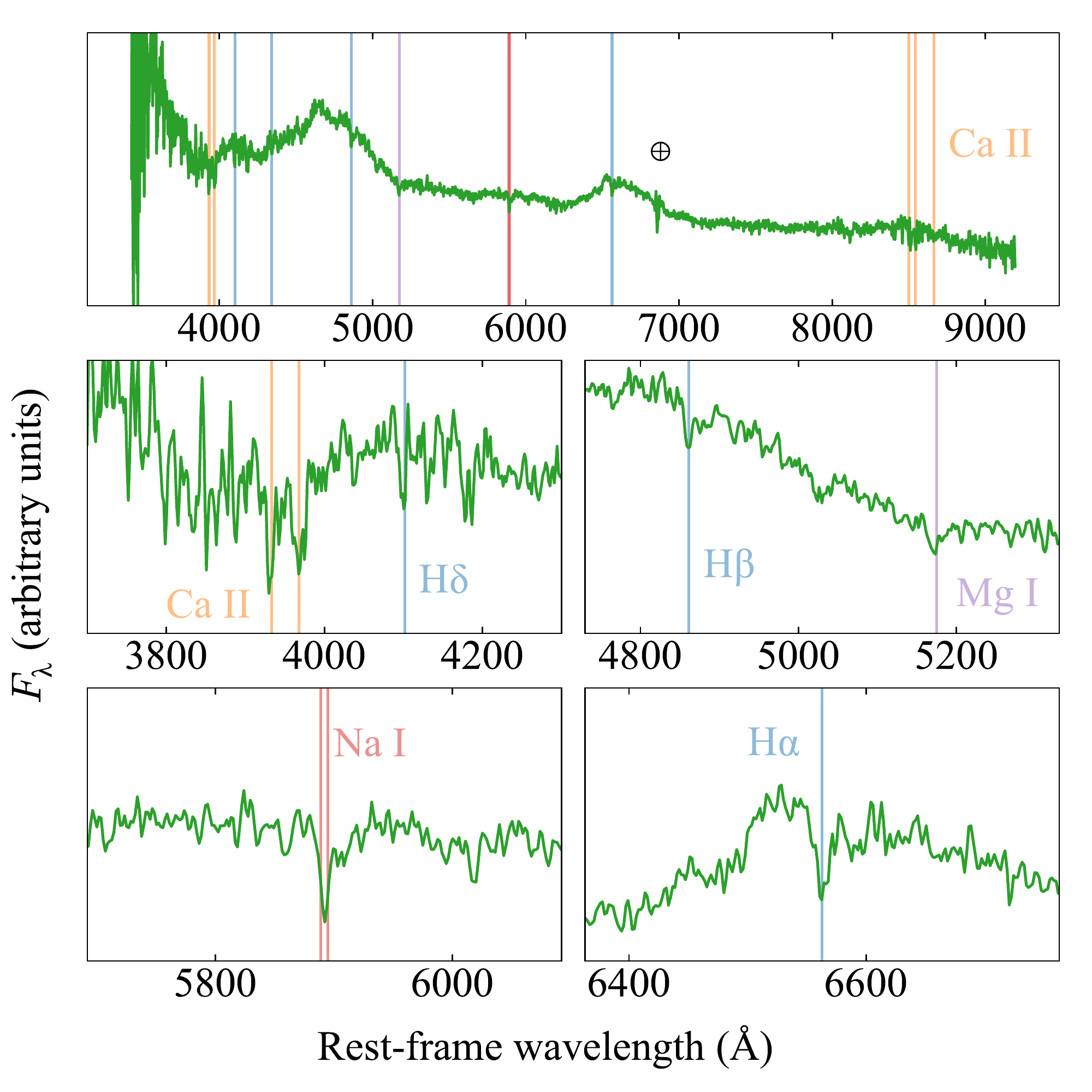}
    \caption{Spectrum of AT2017eqx obtained with Magellan and LDSS3. The blue continuum and broad emission lines, leading to classification as a TDE, can be easily seen. Lower panels show zoom-ins around host galaxy absorption lines from Ca II H\&K, H$\delta$, H$\gamma$, H$\beta$, Mg I 5175, Na I D, H$\alpha$, Ca II NIR triplet, from which we securely measure a redshift $z=0.1089$.}
    \label{fig:z}
\end{figure}

We observed AT2017eqx spectroscopically, beginning on 28-06-2017, as part of a search for nuclear transients. The spectrum with the highest signal-to-noise ratio is shown in Figure \ref{fig:z}. Our observations show the hallmarks of TDEs: a blue continuum, with a roughly constant blackbody temperature of (2--3)$\times 10^4$\,K, and emission lines with widths $>10^4$\,\kms. We also detect narrow absorption lines from the host galaxy, from which we measure the redshift $z=0.1089\pm0.0005$. The host is a Balmer-strong absorption galaxy -- a rare type of galaxy that is greatly over-represented among TDE hosts \citep{Arcavi2014,French2016,LawSmith2017,Graur2018}.

The field was frequently monitored by the Asteroid Terrestrial-impact Last Alert System \citep{Tonry2018}. The ATLAS automated search pipeline triggers on $5\sigma$ detections \citep{Stalder2017,Tonry2018} and no new source (with multiple $5\sigma$ detections) was found by the pipeline. We applied forced photometry on template-subtracted images at the transient coordinates, and manually binned the resulting magnitudes to a nightly cadence. This resulted in several 3$\sigma$ detections in the ATLAS $o$-band, including one prior to the PSST discovery. Synthetic photometry on our earliest spectrum of the transient indicates a colour $o-i=-0.02$\,mag at the time of discovery, in some tension with the PanSTARRS and ATLAS photometry ($o-i=0.29$\,mag). While not entirely explained, this may be related to a broad emission feature at the blue edge of the $i$-band. To compensate, we added a constant shift of 0.31\,mag to all ATLAS measurements so that the colour is consistent with the higher signal-to-noise ratio PanSTARRS photometry.

\section{Follow-up observations}\label{sec:obs}

\subsection{Photometry}

We imaged AT2017eqx in the optical $g,r,i,z$ filters using the Inamori Magellan Areal Camera and Spectrograph (IMACS) \citep{Dressler2011} and the Low Dispersion Survey Spectrograph 3 (LDSS3) on the 6.5-m Magellan Baade and Clay telescopes at Las Campanas Observatory, and KeplerCam on the 1.2-m telescope at Fred Lawrence Whipple Observatory (FLWO).
All images were reduced using \textsc{pyraf} to apply bias subtraction and flat-fielding. Photometry was measured using a custom wrapper for \textsc{daophot}, using stars in the field from PanSTARRS Data Release 1 \citep{Flewelling2016} to determine the point-spread function (PSF) and photometric zeropoint of each image. We downloaded Pan-STARRS1 DR1 $g,r,i$ images \citep{Flewelling2016} as templates for the field and convolved and subtracted them from the images using the \textsc{hotpants} algorithm \citep{Becker2015} to isolate the transient, before measuring its flux with the PSF model. 

We also observed AT2017eqx in $g,r,i$ with the 1-m Swope Telescope at Las Campanas Observatory. The data were reduced using the \textsc{photpipe} photometry and difference imaging pipeline \citep{Rest2005,Kilpatrick2018}. \textsc{photpipe} is a flexible software package that performs optimal bias-subtraction and flat-fielding, image stitching, astrometry, and photometry using \textsc{dophot} \citep{Schechter1993}. We again subtracted PS1 reference images using \textsc{hotpants}. Final photometry was performed on the difference images using \textsc{dophot}.

We obtained further images on 2017-08-16 with the low-resolution imaging spectrograph (LRIS) on the Keck-I 10-m telescope on Mauna Kea, Hawaii. Observations were performed in the blue and red channels simultaneously with the $B+R$ filters and $V+I$ filters and the D560 dichroic, and reduced using \textsc{photpipe}. For our photometric calibration, we used secondary calibrators in each image with magnitudes derived from SDSS standard stars transformed to the BVRI system \citep{Bilir2011,Alam2015}.  Difference imaging was performed using PS1 $g$-band images for the $B$- and $V$-band images, $r$-band images for the $R$-band images, and $i$-band for the $I$-band images.

\begin{figure}
	\includegraphics[width=\columnwidth]{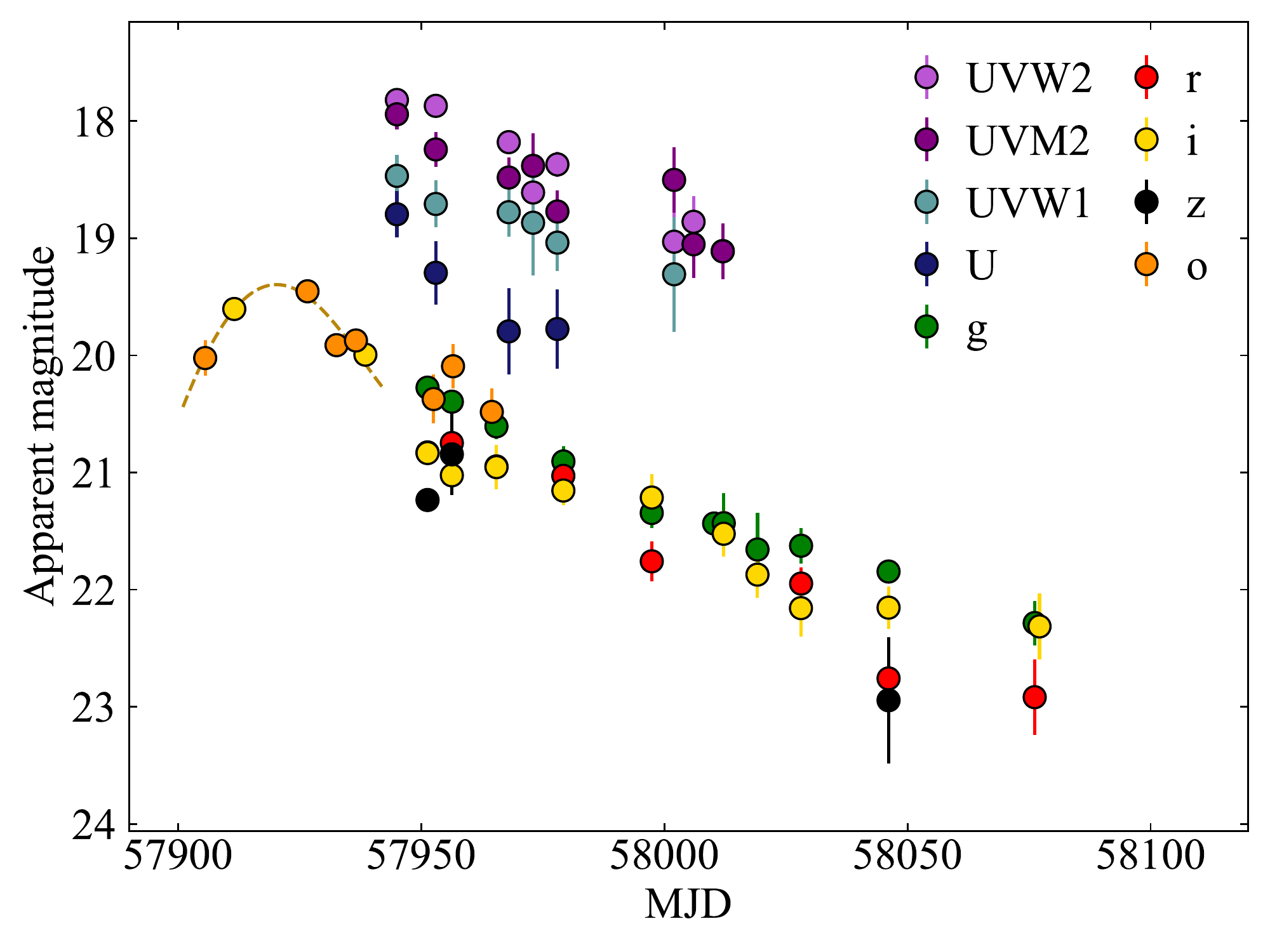}
    \caption{Optical and UV light curves of AT2017eqx from PS1, ATLAS, Magellan, FLWO, Swope and \textit{Swift}. Host fluxes have been removed by subtracting reference images where possible ($g,r,i,z$ and $o$ bands), or otherwise by subtracting fluxes derived from a host SED model (section \ref{sec:host}). A third-order polynomial fit to the early ATLAS and PS1 data (dashed line) suggest maximum light occured on MJD 57921.6.}
    \label{fig:lc}
\end{figure}

Imaging in the UV was obtained using the UV-Optical Telscope (UVOT) on board the Neil Gehrels \textit{Swift} observatory. We downloaded the data from the \textit{Swift} public archive and extracted light curves in the \textit{UVW2}, \textit{UVM2}, \textit{UVW1} and \textit{U} filters following the procedures outlined by \citet{Brown2009}, using a 5'' aperture. The magnitudes are calibrated in Vegamags in the \textit{Swift} photometric system \citep{Breeveld2011}. No reference images were available in the UV, so we estimated the host galaxy contribution using our best-fit spectral energy distribution (SED) model (section \ref{sec:host}). The galaxy flux in each UVOT filter was determined by applying the \textsc{s3} synthetic photometry package \citep{Inserra2018} to the SED model; this was then subtracted from the UVOT measurements of AT2017eqx. Our optical and UV photometry is shown in Figure \ref{fig:lc} and listed in Table \ref{tab:phot}.

\begin{figure*}
\centering
	\includegraphics[width=12cm]{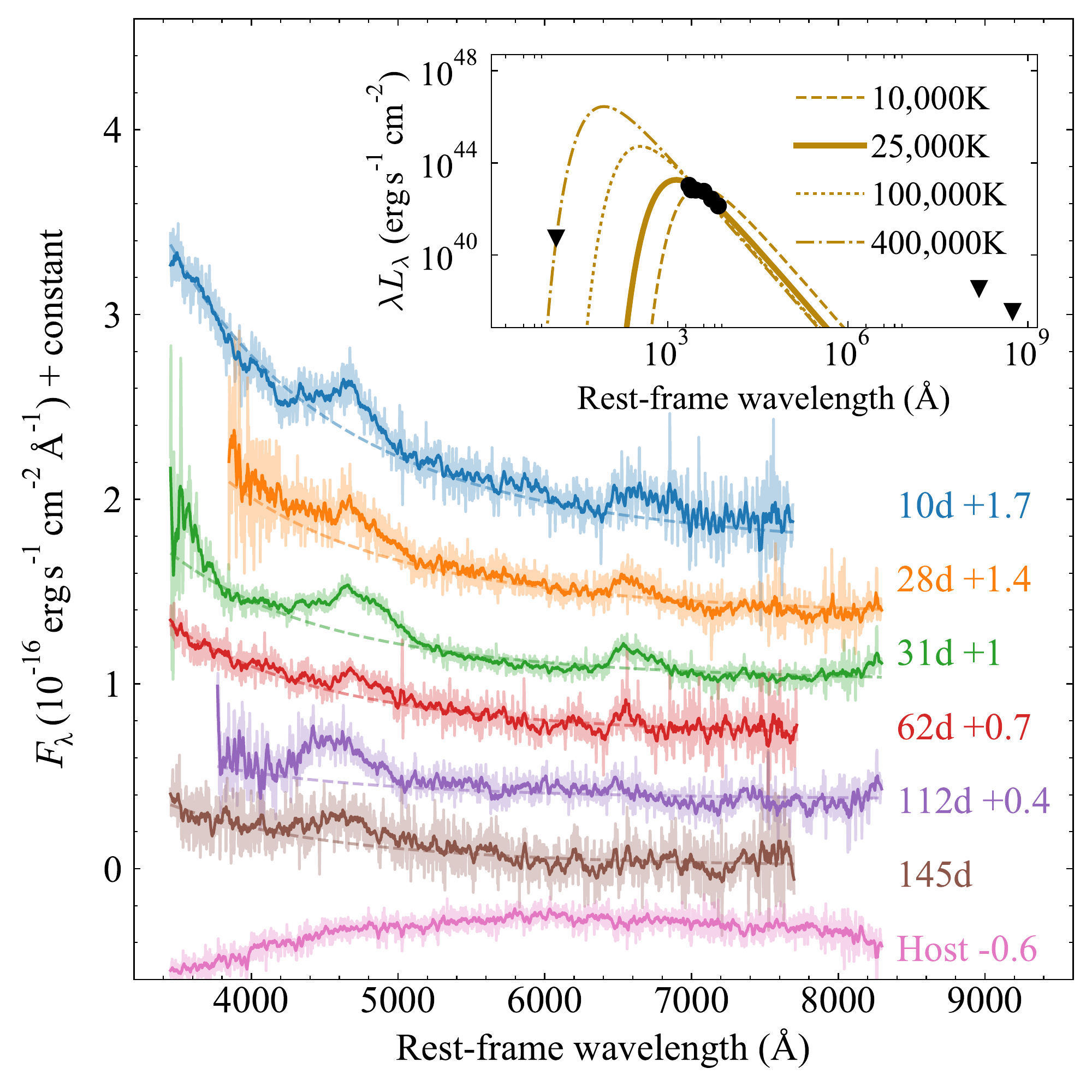}
    \caption{Rest-frame spectra of AT2017eqx, labelled by phase from maximum light. The final spectrum contains only host galaxy light, and has been subtracted from the others. Darker lines show data after Savitsky-Golay smoothing. Dashed lines indicate blackbody fits with temperatures of $\approx 25,000$\,K. Broad emission features from He\,II\,$\lambda4686$ and H$\alpha$ are also apparent, with the latter disappearing in later spectra. The inset shows the full X-ray to radio spectral energy distribution from Chandra and the VLA around day 50.
}
    \label{fig:spec}
\end{figure*}

Finally, we checked for variability in public data from the Wide-field Infrared Survey Explorer \citep[WISE;][]{Wright2010}. The latest data release includes detections in the W1 and W2 bands. These are consistent with the historical magnitudes of the host galaxy. If there is significant dust in the nucleus of this galaxy, future WISE data may show an infrared echo over the next few years \citep{JiangN2016}.

\subsection{Spectroscopy}

We obtained six epochs of spectroscopy between 2017-06-28 and 2017-11-26 using LDSS3, IMACS, and the BlueChannel spectrograph on the 6.5-m MMT telescope \citep{Schmidt1989}. Spectra were reduced in \textsc{pyraf}, including bias subtraction, flat-fielding, wavelength calibration using arc lamps, and flux calibration using standard stars observed on the same nights.
We obtained one additional spectrum on 2018-08-06 using Binospec on MMT. This was reduced using a dedicated pipeline. The final spectrum shows no evidence of TDE features, and we therefore consider it a pure host galaxy spectrum. All spectra were scaled to contemporaneous photometry (interpolated were necessary) and telluric features removed using model fits. We corrected for extinction using the Galactic dust maps of \citet{Schlafly2011}, and assumed negligible extinction in the transient host galaxy. Host-subtracted spectra are plotted in Figure \ref{fig:spec}; a log of spectra is provided in Table \ref{tab:spec}.

\subsection{Radio observations}

We observed AT2017eqx using the Karl G.~Jansky Very Large Array (VLA) in C configuration on 2017-07-14 and 2017-08-22 (Program ID: 16B-318; PI: Alexander). Each observation lasted one hour and was split between C and K bands (central frequencies 6.0\,GHz and 21.7\,GHz). We reduced the data using the VLA CASA Calibration Pipeline (CASA version 4.7.2) and imaged the data using standard CASA routines \citep{McMullin2007}. Both observations used 3C48 as the flux calibrator and J2232+1143 as the phase calibrator. 

\begin{figure}
\centering
	\includegraphics[width=\columnwidth]{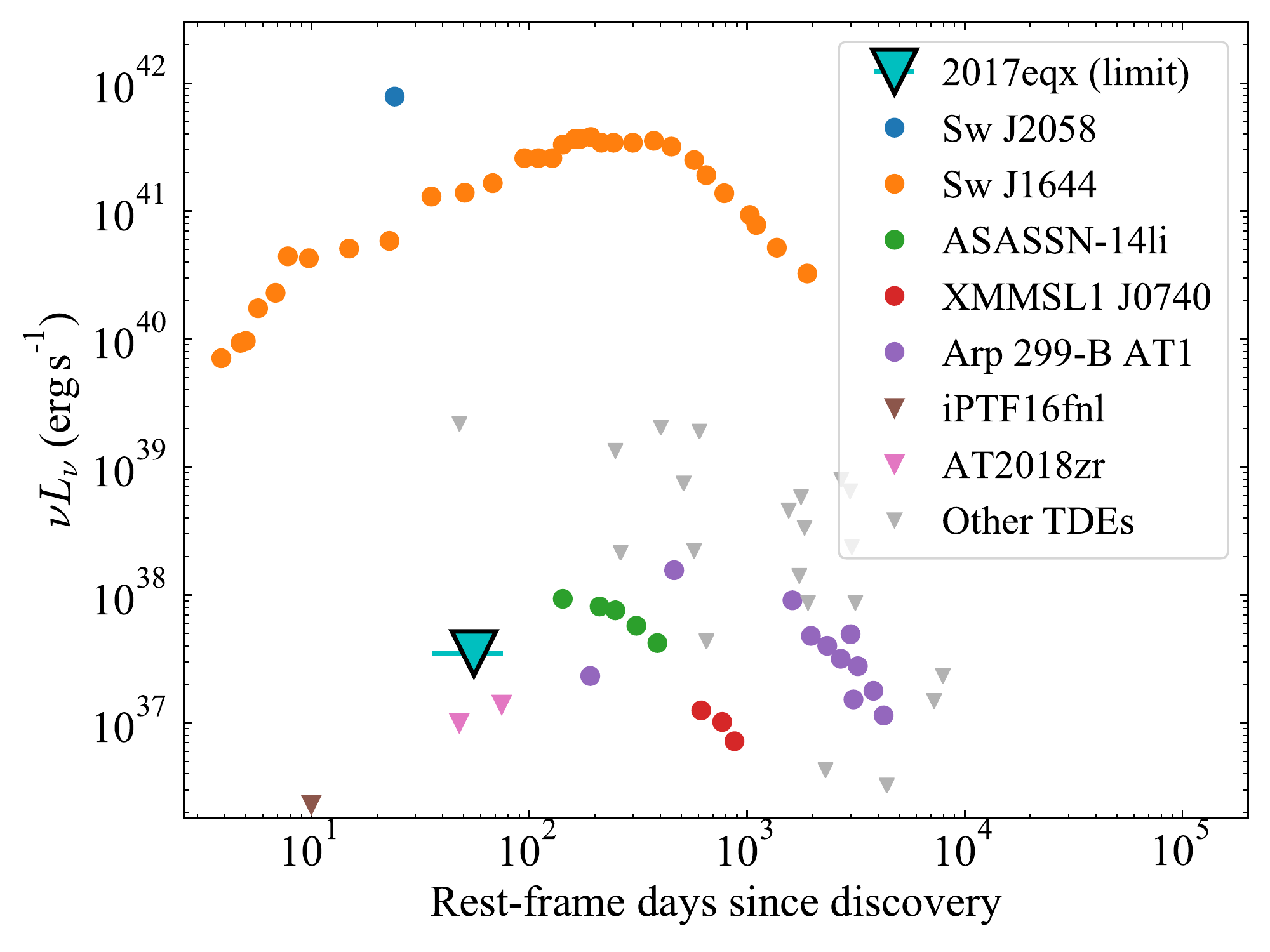}
    \caption{Limit on radio emission at 6\,GHz obtained with the VLA. The deep limit rules out a powerful jet in our line of sight, as seen in the brightest radio TDEs. The possibility of a weaker non-relativistic outflow cannot be excluded, but the limit places strict requirements that the radio luminosity from any an outflow does not exceed that seen in nearby events like ASASSN-14li \citep{Alexander2016,vanVelzen2016}. Further observations will be required to rule out an off-axis jet such as that seen in Arp 299-B \citep{Mattila2018}. Comparison data are from \citet{Alexander2017,Eftekhari2018,Blagorodnova2017,Zauderer2011,Berger2012,Cenko2012,Alexander2016,Alexander2017,Eftekhari2018,Komossa2002,Bower2013,vanVelzen2013,Arcavi2014,Chornock2014,Mattila2018,vanVelzen2018}}
    \label{fig:radio}.
\end{figure}

No significant radio emission was detected in either epoch. We therefore combined the epochs to derive deeper limits on the radio flux, at a mean phase of 55 days after discovery. The 6\,GHz limit corresponds to $\nu L_\nu < 3.5\times10^{37}$\,\ergs at the distance of AT2017eqx. All limits are listed in Table \ref{tab:radio}. AT2017eqx has one of the deepest radio limits among TDEs, particularly at this phase from disruption (Figure \ref{fig:radio}). Only iPTF16fnl \citep{Blagorodnova2017} and AT2018zr \citep{vanVelzen2018} have deeper constraints.

The non-detection in the radio rules out a powerful jet similar to those associated with the relativistic TDEs Swift\,J1644+57 \citep{Zauderer2011,Bloom2011,Burrows2011} and Swift\,J2058+05 \citep{Cenko2012}. However, most TDEs are not bright in the radio. Simulations by \citet{Mimica2015} suggested that this cannot be explained solely by beaming effects, implying that jetted TDEs are intrinsically rare. ASASSN-14li was detectable at radio wavelengths due to its proximity (90\,Mpc), with a peak luminosity 3-4 orders of magnitude lower than the {Swift} events. This modest radio emission was interpreted as a weak jet \citep{vanVelzen2016} or a wide-angle non-relativistic outflow \citep{Alexander2016}. A similar outflow was also detected in XMMLS1\,J0740-85 \citep{Alexander2017}, another very nearby TDE (at 75\,Mpc). Despite the much greater distance to AT2017eqx (520\,Mpc), our VLA limits are sufficiently deep to rule out emission at 6\,GHz in excess of that seen in ASASSN-14li, though slightly fainter emission comparable to XMMSL1\,J0740-85 is still permitted.

Future observations at similar sensitivity may determine if AT2017eqx could have launched an off-axis jet that later spreads into our line of sight. The off-axis TDE jet resolved in Arp 299-B AT1 by \citet{Mattila2018} is an example of such a system. Regardless, our limit is among the deepest observational constraints on a non-relativistic outflow during the first months after disruption for any TDE to date.

\subsection{X-ray observations}

\begin{figure}
\centering
	\includegraphics[width=\columnwidth]{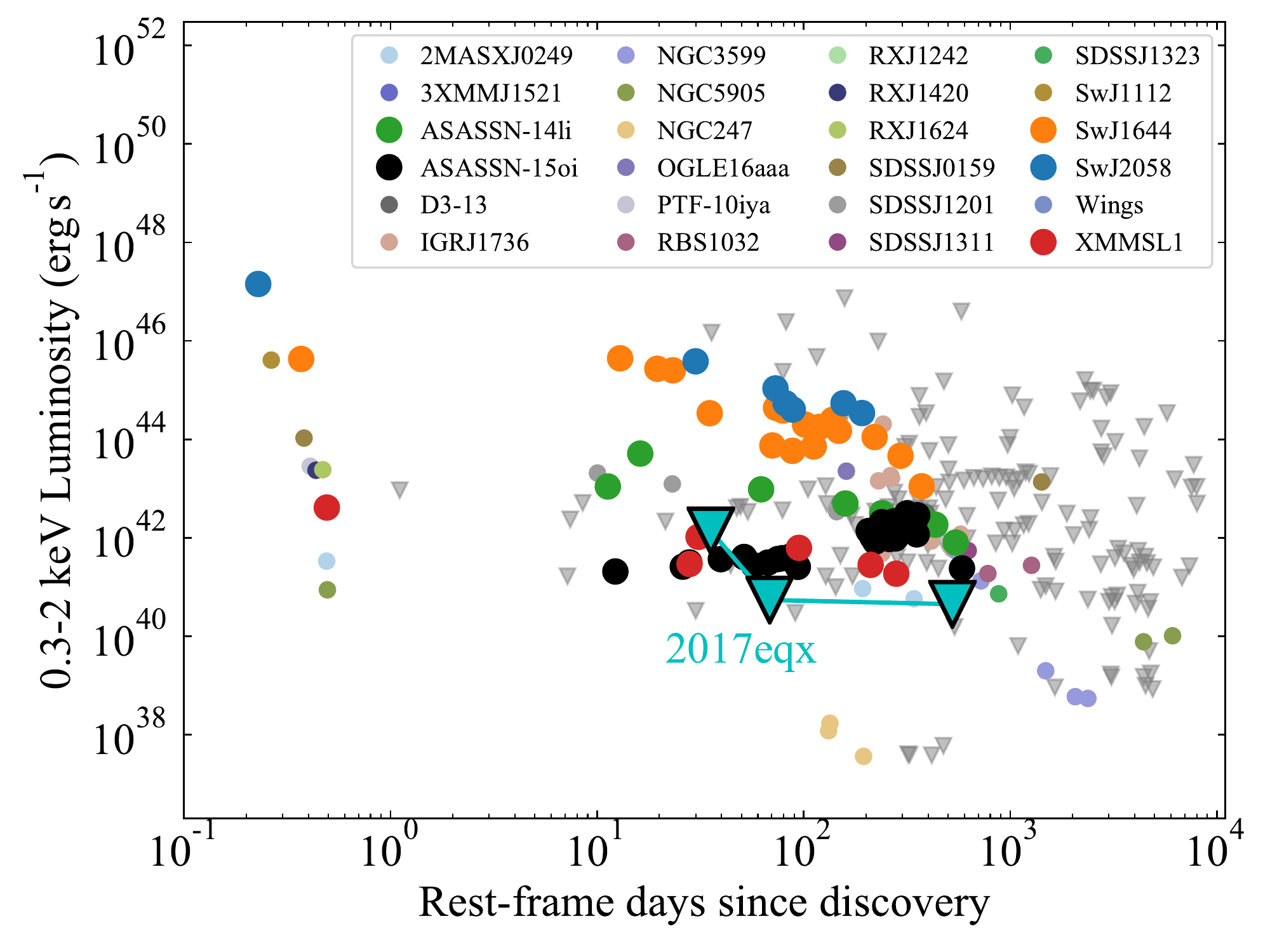}
    \caption{Limits on X-ray emission obtained using \textit{Swift} and \textit{Chandra}. The limits imply an X-ray to optical ratio $< 10^{-2}$, much lower than for any TDEs that do have observed X-ray emission. Comparison data are from \citet{Auchettl2017,Gezari2017}. TDEs plotted with large symbols use the same colour scheme as in Figure \ref{fig:radio}.}
    \label{fig:x}
\end{figure}

Initial X-ray data was obtained using the X-ray Telescope (XRT) onboard \textit{Swift}. Analysis of 5.7\,ks of data collected between 2017-07-11 and 2017-07-24 revealed no significant flux in the 0.3-10\,keV range. 
We carried out deeper observations using the \textit{Chandra} X-ray Observatory with 10\,ks integrations on 2017-08-16 and 2019-01-04 (Programs 20625, 21437; PI: Nicholl, Berger), again resulting in non-detections. 
We also checked for archival imaging before the optical flare to rule out previous AGN activity. This field was observed by XMM-Newton on 2015-05-20 (Observation ID: 076247020). We analysed the image using the online XMM-Newton Science Archive tools. No source is detected at the position of the host galaxy.

For all epochs, we assume a power law spectral model with $\Gamma=2$, and a Milky Way hydrogen column density of $N_H=5.06\times10^{20}$\,cm$^{-2}$ along this line of sight, in order to convert count rates to fluxes over the range 0.3-10\,keV. We verified that our results are only weakly dependent on our choice of model, and we derive similar constraints if we instead assume a blackbody SED with a temperature of 0.1\,keV. The limiting count rates and fluxes are given in Table \ref{tab:x}.

Our limits from \textit{Chandra} imply an X-ray/optical ratio $< 10^{-2}$. TDEs with optical and X-ray detections have generally exhibited X-ray/optical ratios $\sim 1$. For some TDEs, the ratio can be much greater. These are generally relativistic TDEs, though XMMSL1 J074008.2-853927 \citep{Saxton2017} appears to be an example of a TDE with a thermal component that also exhibited a large X-ray/optical ratio. However, many TDEs have X-ray non-detections that imply a ratio $\ll 1$ \citep{Auchettl2017}. AT2017eqx falls firmly in this category. Compared to other TDEs which have exhibited strong X-rays, e.g.~ASASSN-14li \citep{Miller2015,Brown2017}, ASASSN-15oi \citep{Holoien2016b,Gezari2017,Holoien2018b}, and Swift\,J1644+57 \citep{Levan2011,Levan2016}, the X-ray emission from AT2017eqx is at least an order of magnitude less than that observed in those events (Figure \ref{fig:x}).

\section{Photometric analysis}\label{sec:phot}

\subsection{Bolometric luminosity and temperature}

\begin{figure}
\centering
	\includegraphics[width=\columnwidth]{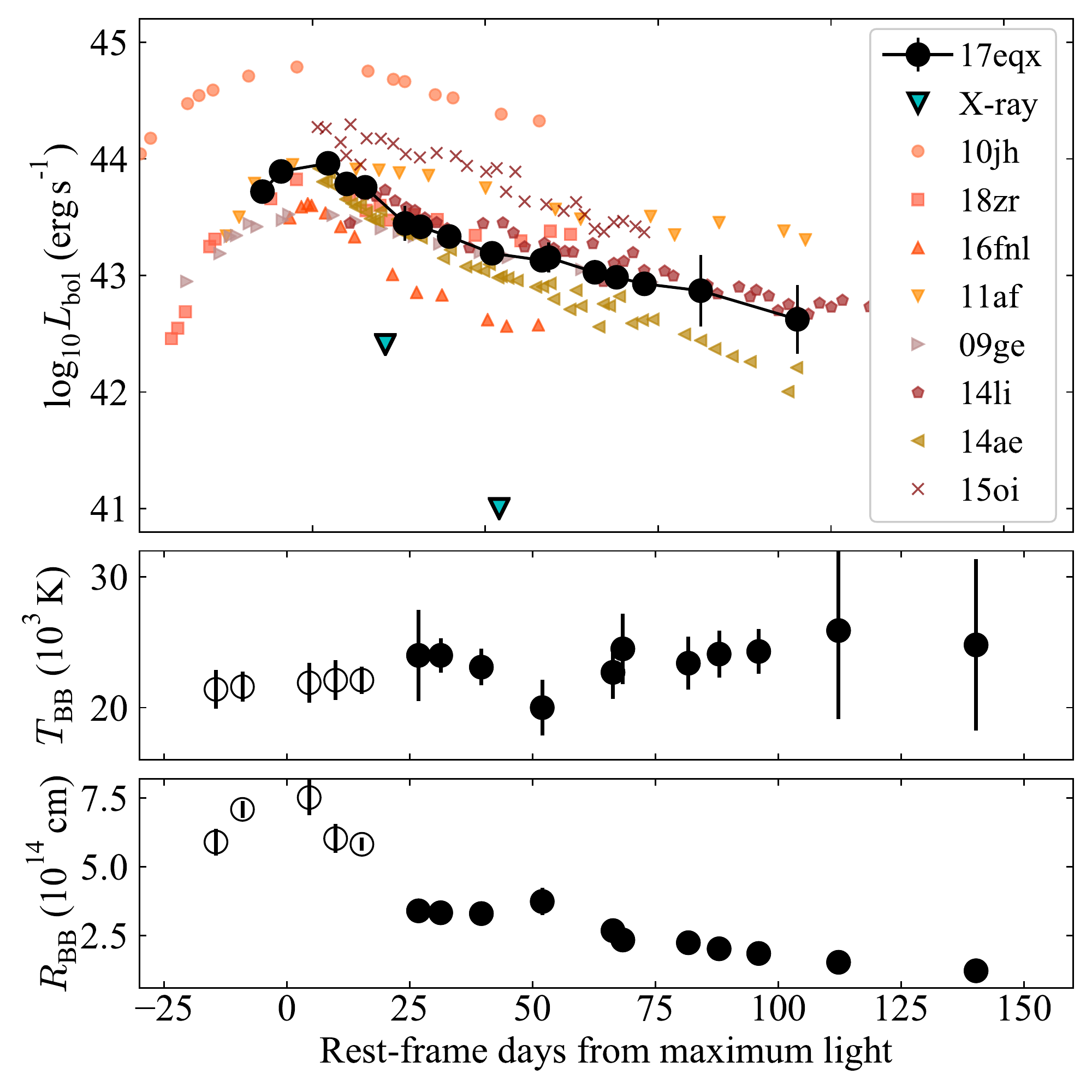}
    \caption{
    Bolometric light curve of AT2017eqx using blackbody fits, calculated at all epochs with $i$ or $o$ observations. Empty symbols indicate early epochs where the temperature was estimated assuming zero colour evolution. We also show the deep X-ray limits corresponding to $\lesssim 1\%$ of the optical luminosity. 
 The light curve evolution is similar to other TDEs from the literature \citep{vanVelzen2018,Holoien2016b,Gezari2012,Chornock2014,Arcavi2014,Holoien2014,Holoien2016,Blagorodnova2017}.
    The middle and lower panels show the best-fit temperature and radius for each epoch. The emitting material contracts at roughly constant temperature. }
    \label{fig:bol}
\end{figure}

We determine the bolometric luminosity of AT2017eqx using the following procedure, implemented via \textsc{superbol} \citep{Nicholl2018}. We first interpolate or extrapolate all light curves to a common set of observation times, defined by having an observation in $i$ or $o$ bands, using low (first- or second-) order polynomial fits. At each epoch, we integrate the spectral luminosity over all bands, and fit blackbody curves to estimate the flux falling outside of the UV-optical wavelength range. These fits also allow us to constrain the temperature and radius of the emitting material.

The bolometric light curve and blackbody parameters are shown in Figure \ref{fig:bol}.
The best-fitting blackbody temperature is $\approx 21,000-25,000$\,K and shows no significant evolution in time within the errors. Visually there may be a slight rise, though this is sensitive to extrapolating the UV photometry. The blackbody radius decreases from $7.6\times10^{14}$\,cm to $1.2\times10^{14}$\,cm. AT2017eqx peaks at a luminosity $\approx 10^{44}$\,\ergs\ (with a solid lower limit of $>10^{43.5}$\,\ergs\ in the observed bands), and emits a total of $> 4.4\times10^{50}$\,erg over the duration of our observations. The luminosity and light curve shape is typical of optical TDEs.

\subsection{TDE model fit}

We fit a physical TDE model to the observed UV and optical photometry using \textsc{mosfit}: the Modular Open Source Fitter for Transients \citep{Guillochon2018}. This is a semi-analytic code employing a range of modules that can be linked together to produce model light curves of astronomical transients, and determine the best fitting model parameters through Bayesian analysis. The TDE model and associated modules in \textsc{mosfit} are described in detail by \citet{Mockler2018}. The method is based on an older code, \textsc{tdefit} \citep{Guillochon2014}, and uses a combination of scaling relations and interpolations between the output of numerical TDE simulations to determine the luminosity.

The model has ten free parameters: the masses of the star and SMBH; the impact parameter (determining whether a disruption is full or partial); the efficiency of converting fallback energy into radiation; two parameters controlling the relationship between the luminosity and radius; the time of disruption relative to first detection; a viscous timescale over which the accretion disk forms; the extinction (column density) in the host galaxy; and a white-noise term parameterising any unaccounted-for variance. 

\begin{figure}
\centering
	\includegraphics[width=\columnwidth]{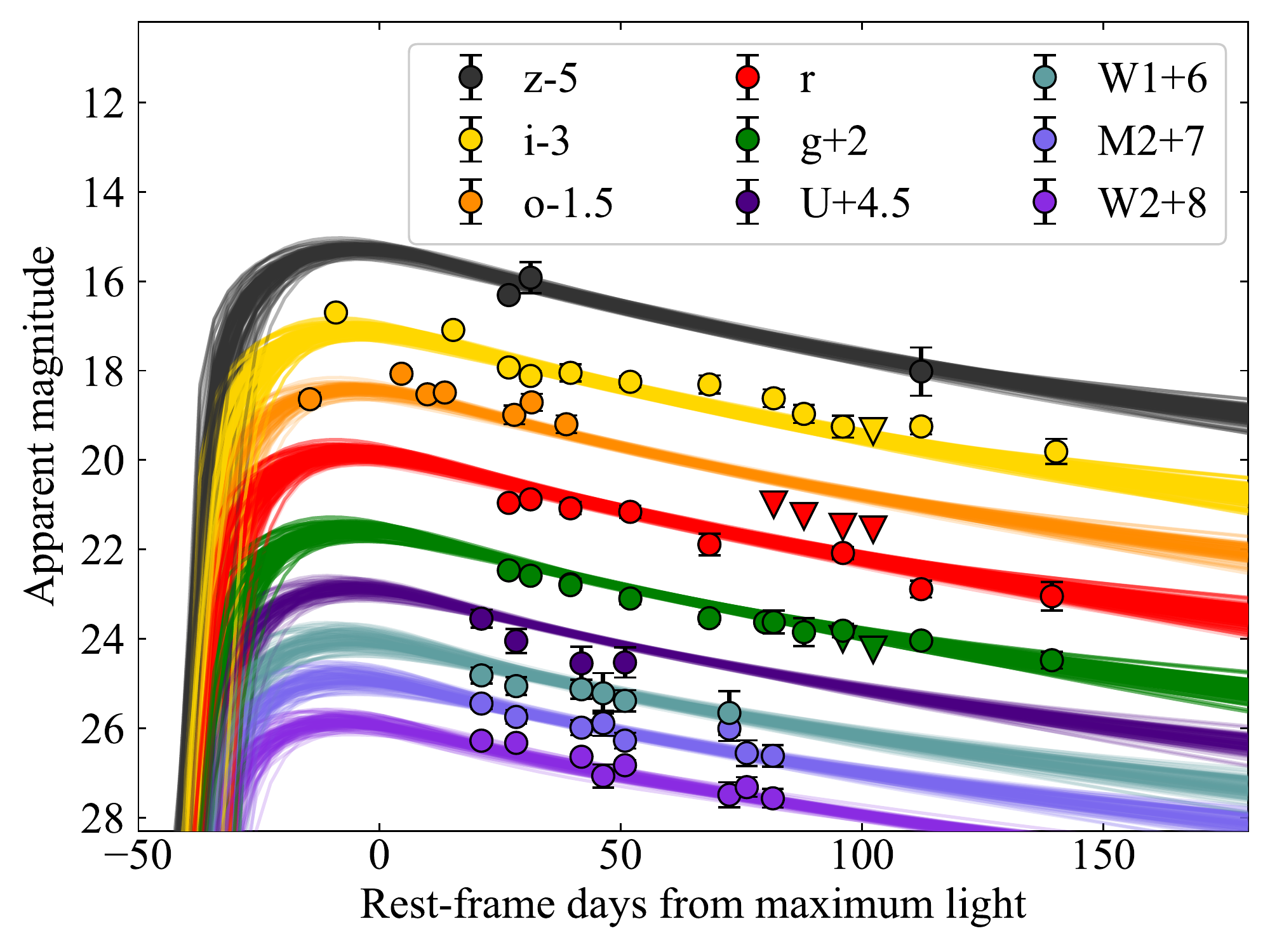}
    \caption{Fits to the light curve of AT2017eqx using the TDE model in \textsc{mosfit}, for the case where we assume disruption occured within 30 days before the first detection. Coloured lines show 100 MCMC realisations.}
    \label{fig:fit}
\end{figure}

\begin{figure*}
\centering
	\includegraphics[width=\textwidth]{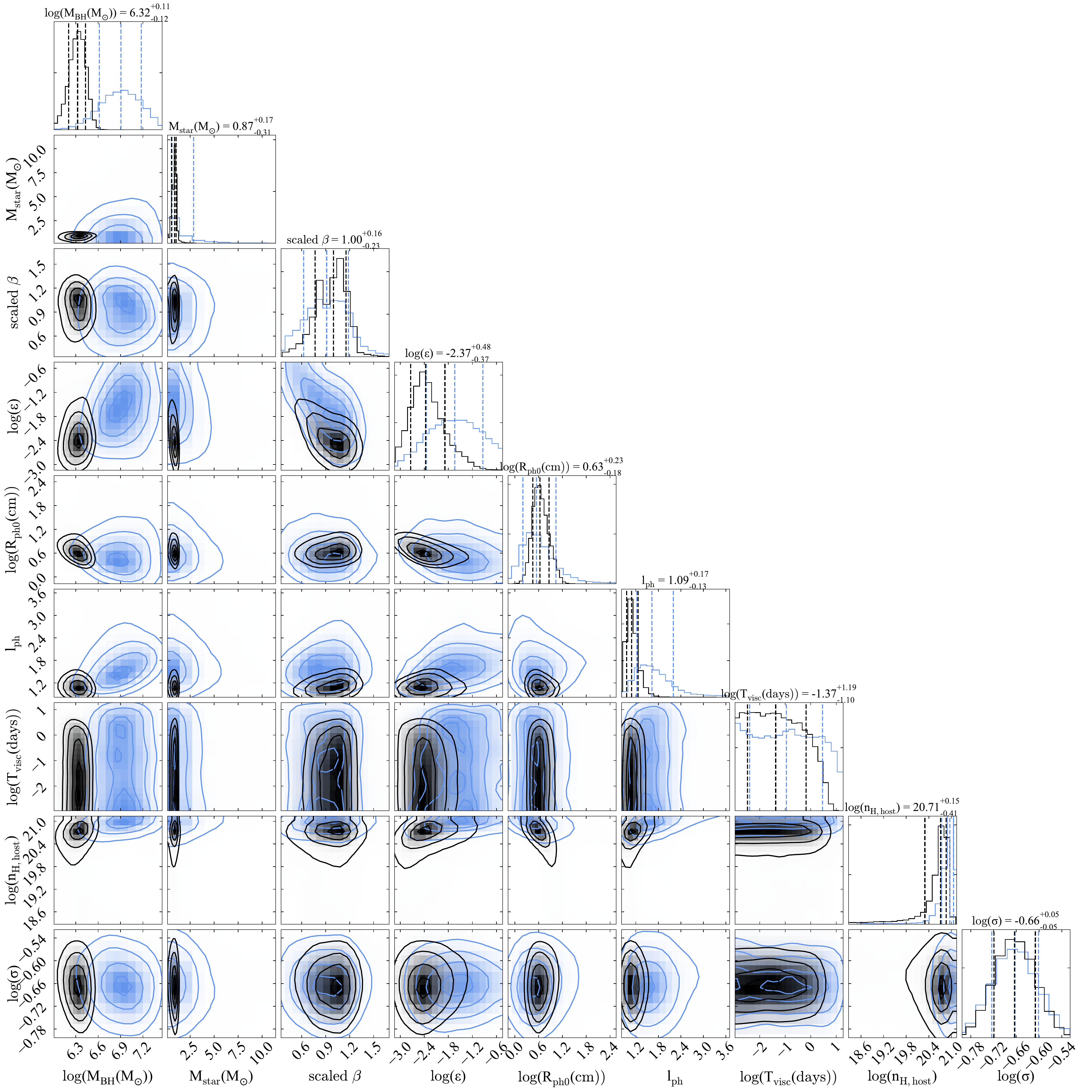}
    \caption{Corner plot showing the posteriors for our TDE model fit. Black corresponds to the solution with a tight prior on the disruption time ($<30$ days before detection); blue corresponds to the solution with a broader prior using the last non-detection in ATLAS ($<168$ days). 
}
    \label{fig:corner}
\end{figure*}

To sample the parameter space we used the affine-invariant ensemble method \citep{Goodman2010,Foreman2013} as implemented in \textsc{mosfit}. We ran the Markov Chain with 100 walkers for 50,000 iterations, checking for convergence by ensuring that the Potential Scale Reduction Factor was $<1.2$ at the end of the run \citep{Brooks1998}. We assume the same priors as used by \citet{Mockler2018}, with the exception of the time of disruption. Given that we have only one or two data points weakly constraining the rise-time of AT2017eqx, we run two fits: one where we allow the disruption to occur up to 168 days before first detection (corresponding to the time of the last ATLAS non-detection); and one where we restrict disruption to within 30 days before detection, to force a better fit to the first ATLAS detection.

The model light curves for the latter case are shown in Figure \ref{fig:fit}, and the posteriors of the parameters for both fits are shown as a corner plot in Figure \ref{fig:corner}. These posteriors are overall similar, with narrower constraints in the case with the stronger prior on disruption time, though some posteriors show shifts in the median of up to $\sim2\sigma$ between the fits.

The posteriors point to the full disruption (i.e.~scaled impact parameter $\approx 1$) of a  $\sim0.9$\,\M\ star. The SMBH mass is more sensitive to the assumed rise time, with $\log(M_{\rm BH}/M_\odot)=6.9\pm0.3$ in the more general case, falling to $\log(M_{\rm BH}/M_\odot)=6.3\pm0.1$ when the rise time is constrained to 30 days. The TDE peak luminosity ($\approx 10^{44}$\,\ergs) corresponds to 10-40\% of the Eddington luminosity for this SMBH mass range. The model employs a power-law photospheric radius, $R\propto L^l$; the best fit has $l\simeq1$, i.e.~the radius of the emitting region is directly proportional to the luminosity.

The best-fit viscous time is always much shorter than the rise time, indicating that the emission has not been delayed by a long circularisation process. This requires either that an accretion disk forms promptly after disruption or that the optical luminosity instead arises from stream-stream collisions \citep{Mockler2018}. Assuming that the energy is released close to the innermost stable circular orbit of a $10^{6.3}$\,\M\ SMBH, the implied blackbody temperature is $\sim 5\times10^5$\,K, which is in some tension with the temperature range ruled out by our X-ray non-detections, $\gtrsim 4\times10^5$\,K (Figure \ref{fig:spec}, inset). Thus in an accretion-powered model, the disk emission would have to be downgraded to a cooler spectrum by a reprocessing layer.

\section{Spectroscopic analysis}\label{sec:spec}

\begin{figure}
	\includegraphics[width=\columnwidth]{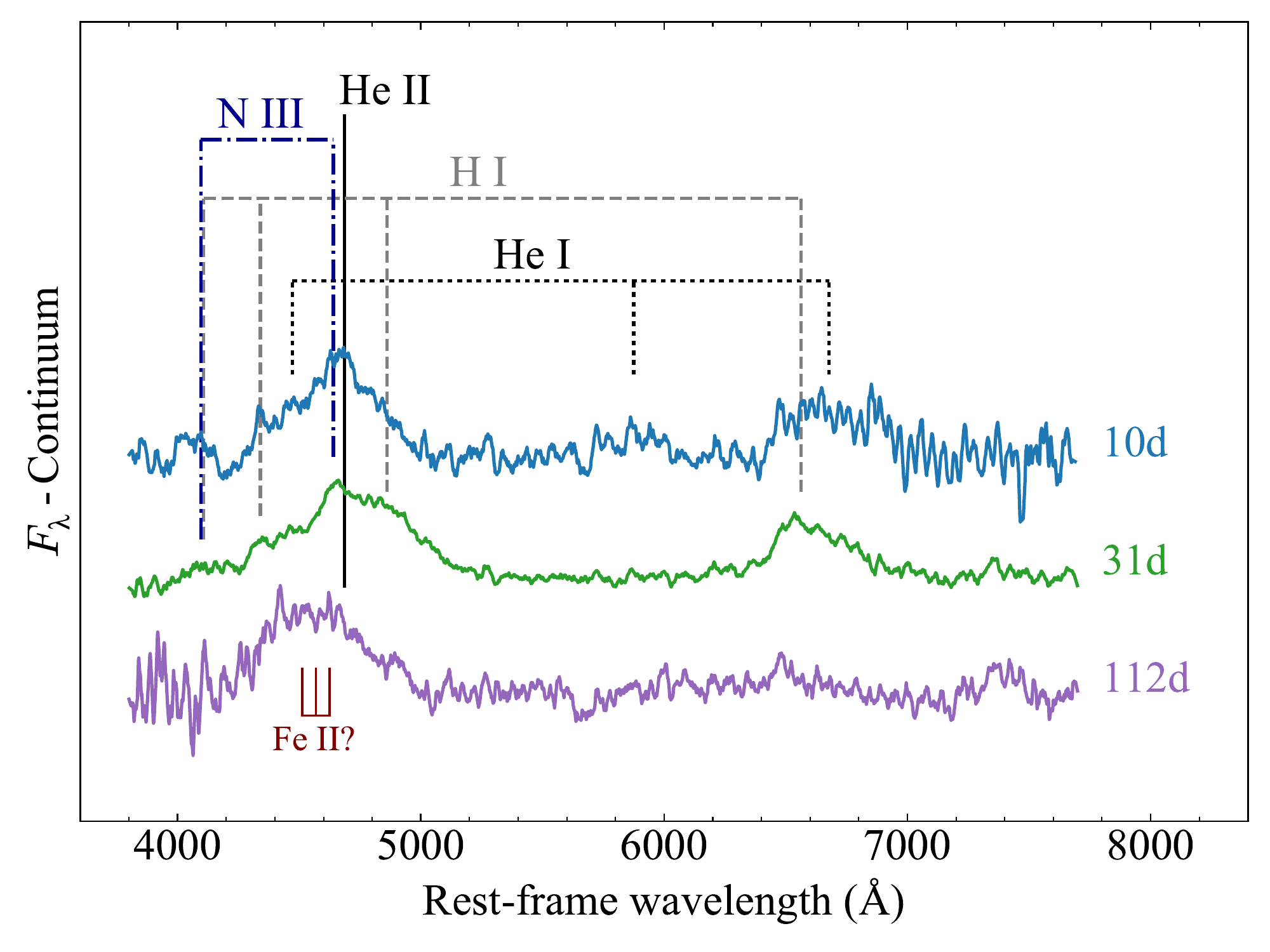}
    \caption{Continuum-subtracted, smoothed spectra of AT2017eqx, with important line features marked. The early spectra show H\,I and He\,II lines, possibly contaminated by He\,I, while later spectra show only a blueshifted feature close to He\,II. The peak of the He\,II line at early times appears to have an additional contribution from N\,III, indicating the Bowen flourescence mechanism may be in effect \citep{Blagorodnova2018,Leloudas2019}. We also mark the location of Fe\,II lines recently identified in AT2018fyk by \citet{Wevers2019b}, however we disfavour this identification for AT2017eqx due to the lack of both X-rays and an optical plateau as seen in that event.}
    \label{fig:lines}
\end{figure}

\subsection{Line identification}

As shown in the previous section, the overall blue continuum in the spectrum shows little change over time. To analyse the evolution in our spectra, we therefore first remove this continuum using fifth-order polynomial fits \citep[e.g.][]{Hung2017}. We excise from the fit any regions within $\pm150$\,\AA\ ($\approx 10,000$\,\kms) of the rest wavelengths of H$\alpha$, H$\beta$ and He\,II $\lambda$4686 -- typically the dominant features in optical TDE spectra.
Figure \ref{fig:lines} shows examples of the smoothed TDE spectra after subtraction of the continua, with line features marked. 

The early spectra exhibit two strong broad emission lines, along with some weaker features. The strongest lines peak close to the rest wavelengths of H$\alpha$ and He\,II $\lambda4686$, with several features matching the locations of the other Balmer lines. The presence of both He\,II and H\,I in the spectrum would qualify AT2017eqx as a transitional event between so-called `H-rich' and `He-rich' TDEs, following the continuum identified by \citet{Arcavi2014}. However, later spectra, $>62$ days after maximum, look markedly different. The hydrogen lines have largely disappeared, leaving only a strong emission line around He\,II $\lambda4686$. AT2017eqx therefore offers direct evidence that disruption of a hydrogen-rich star (required by the early spectra) can form a spectrum with no visible hydrogen. We will return to this critical point in section \ref{sec:interp}.

There is some indication of neutral He\,I lines in the spectrum, which could help to explain asymmetries apparent in the line profiles. A red shoulder in the early-time H$\alpha$ profile could be due to He\,I $\lambda6678$, while a blue shoulder in He\,II $\lambda4686$ could be contamination from He\,I $\lambda4471$. The earliest spectrum shows a potential He\,I $\lambda5876$, which might support this interpretation, though He\,I lines do not appear to be present at later times (see section \ref{sec:profiles}).

While early studies of TDE optical spectra primarily identified emission from hydrogen and helium, several authors have recently found evidence for metal lines such as N\,III and O\,III, attributed to the Bowen flourescence mechanism \citep{Blagorodnova2018,Leloudas2019}\footnote{Lines from other ionisation states of oxygen, nitrogen and carbon have been found in the UV \citep{Cenko2016,Brown2018}.}. This occurs when the $2\rightarrow1$ transition in recombining He\,II emits a photon that happens to resonate with a far-UV transition in O\,III, which in turn produces a photon that resonates with N\,III \citep{Bowen1935}. Optical photons are also produced in this cascade, which (unlike the far-UV photons) can escape to the observer. We mark the positions of N\,III lines on Figure \ref{fig:lines} (the O\,III Bowen lines are at $\lambda<4000$\,\AA). N\,III $\lambda$4100 falls too close to H$\delta$ for a firm identification. However, N\,III $\lambda4641$ (or possibly C\,III $\lambda4649$) does appear to be present in the early spectra of AT2017eqx, manifesting as a bump just bluewards of the centre of the broad He\,II $\lambda4686$ peak. This is similar to the profile seen in TDEs iPTF15af \citep{Blagorodnova2018} and AT2018dyb/ASASSN-18pg \citep{Leloudas2019}. This supports the claim by \citet{Leloudas2019} that such features may be common in TDEs.

The He\,II line appears shifted to bluer wavelengths in the later spectra. The shift is too large to be explained by blending with N\,III $\lambda4641$; furthermore, the N\,III $\lambda$4100 line is not strong like that seen in AT2018dyb, indicating that Bowen features are unlikely to dominate the blend. It is possible that there is a contribution from Fe\,II, which we also mark on the figure. This was recently identified by \citet{Wevers2019b} in the spectrum of AT2018fyk/ASASSN-18ul. In that case the appearance of the line, thought to originate from dense gas close to an accretion disk, was associated with X-ray emission and a plateau in the optical and UV light curves. Other TDEs have exhibited narrow Fe\,II emission, likely from pre-existing broad-line regions around the SMBH \citep{Blanchard2017}, but their spectra were dominated by multi-component Balmer emission. We see none of these properties in the case of AT2017eqx, and therefore cannot identify Fe\,II.

\subsection{Line profiles and velocities}
\label{sec:profiles}

We analyse the line profiles quantitatively by means of Gaussian fits. There are several important caveats to note: blending between broad overlapping lines complicates their observed profiles. The fits are also sensitive to the choice of continuum, which we defined using a fifth-order polynomial fit. Finally, the profile of even an isolated line is not necessarily Gaussian -- lines may have a large electron-scattering optical depth, and outflows can induce asymmetries \citep{Roth2018}. Nevertheless, we proceed with Gaussian fits as a simplified means to determine line centres and widths, following other studies in the literature \citep{Arcavi2014,Blagorodnova2017,Hung2017}.

\begin{figure}
\centering
	\includegraphics[width=\columnwidth]{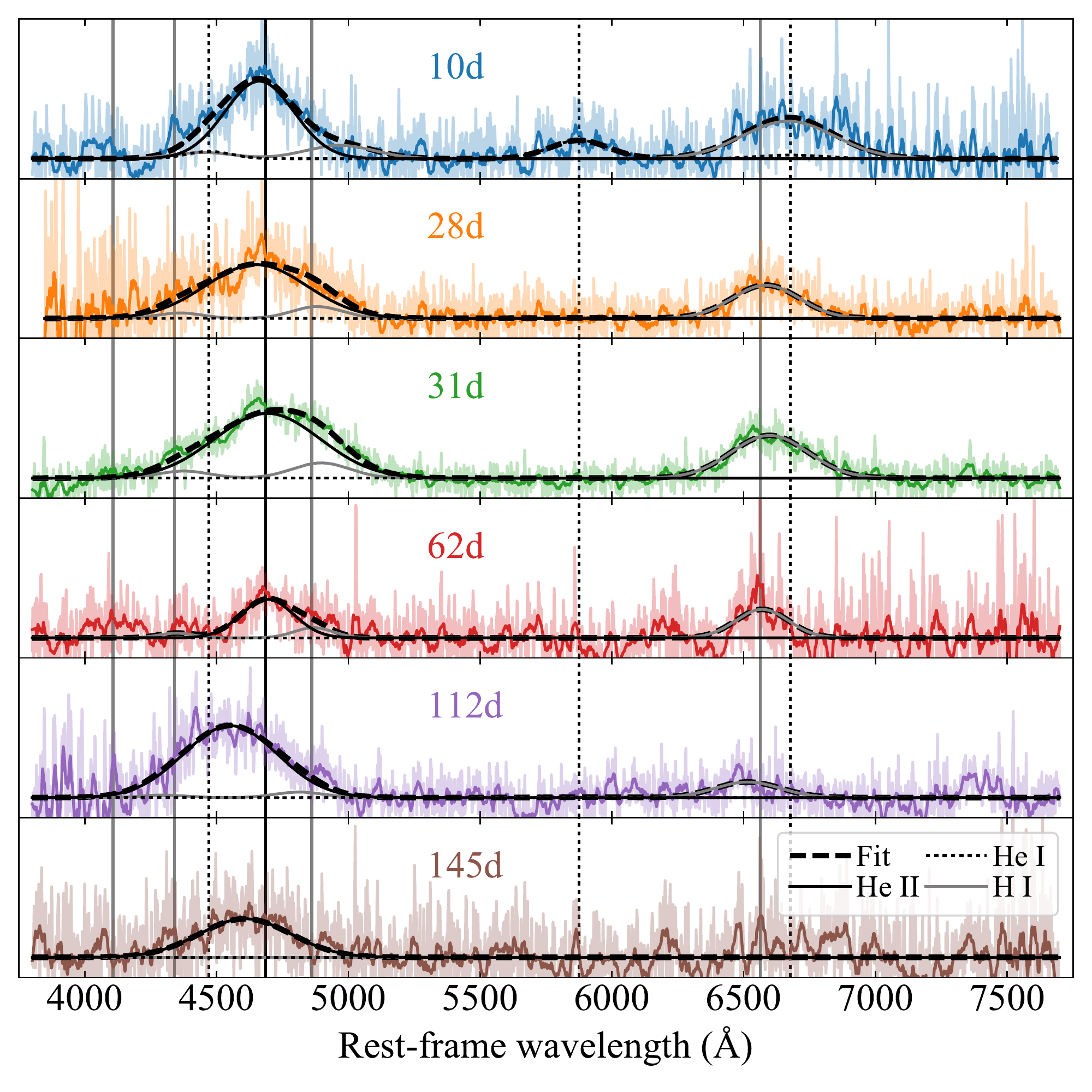}
    \caption{Fits to the main emission features in the spectra of AT2017eqx. 
    Lines of the same ion (H\,I, He\,II, He\,I) have the same velocity centroids and widths as well as fixed luminosity ratios. The fits are to the unsmoothed spectra, but smoothed data are also shown to guide the eye. Derived parameters for H$\alpha$ and He\,II $\lambda4686$ are shown in Figure \ref{fig:linev}.
    }
    \label{fig:profiles}
\end{figure}

\begin{figure}
\centering
	\includegraphics[width=\columnwidth]{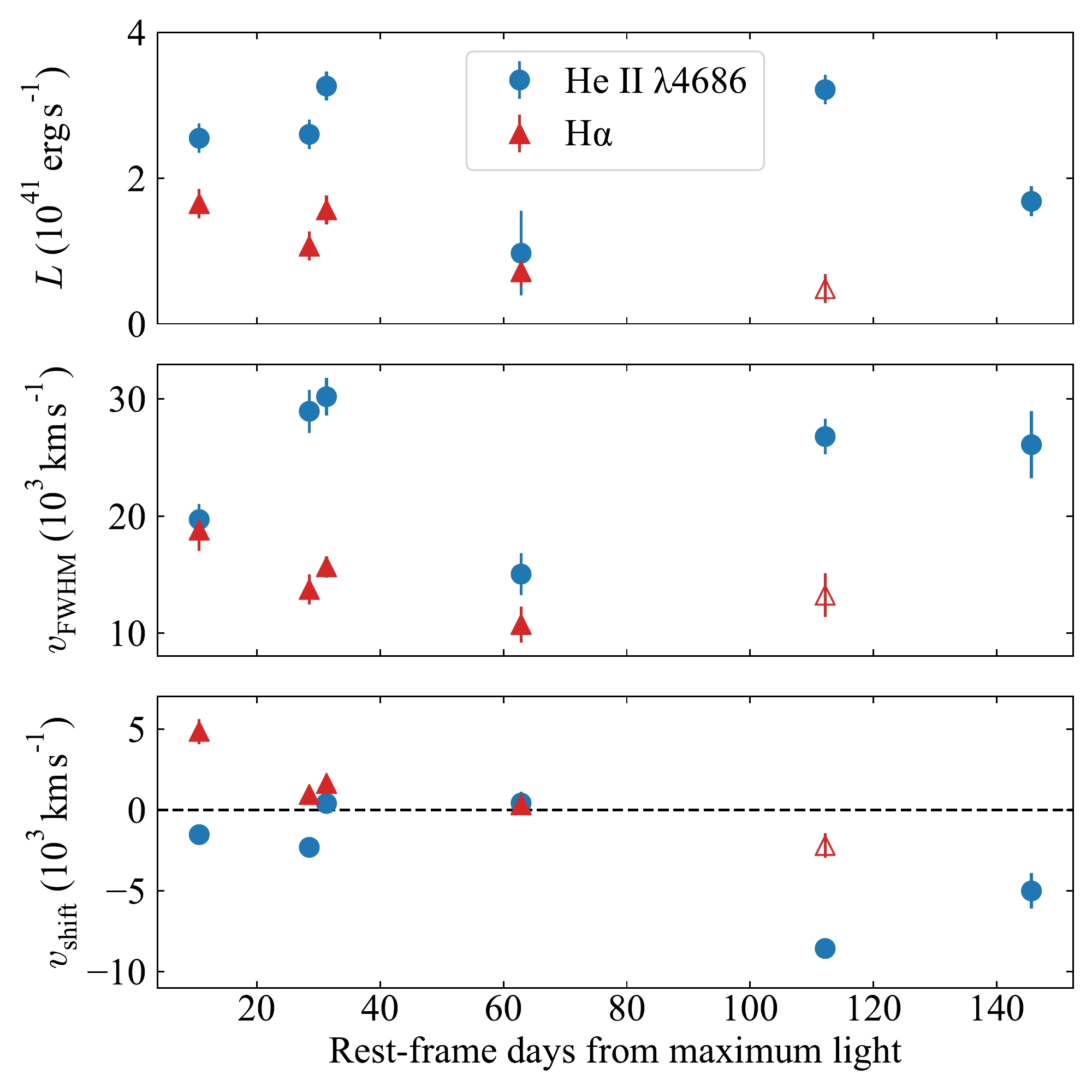}
    \caption{Evolution of line properties from the fits in Figure \ref{fig:profiles}. The He\,II/H$\alpha$ ratios changes from $\sim 2$ to $>10$, while He\,II develops a blueshift of 5000-8000\,\kms. We are unable to reliably measure a blueshift for H$\alpha$ in the later epochs due to its low luminosity (empty symbol).}
    \label{fig:linev}
\end{figure}

The most difficult issue to deal with is the effect of blending. We account for this by fitting the entirety of each continuum-subtracted spectrum simultaneously (Figure \ref{fig:profiles}) with a model that includes H\,I ($\alpha,\beta,\gamma$), He\,II (4686\,\AA) and He\,I (4471, 5876, 6678\,\AA). We neglect the Bowen fluorescence lines (N\,III and C\,III) as these are almost completely degenerate with He\,II $\lambda4686$ at the observed line widths, and we do not see a strong N\,III $\lambda4100$ line to suggest that the Bowen mechanism dominates line formation in this region, as seen in AT2018dyb \citep{Leloudas2019}. We therefore estimate that the Bowen lines account for no more than $\lesssim20\%$ of the measured flux in this linr blend.

For each line we allow the centroid offset (red/blueshift), velocity width and luminosity to vary, but fix lines from the same ion to have the same offset and width. 
To further reduce the number of free parameters, we fix the ratios between lines from a given ion. For the Balmer lines we assume Case B recombination \citep{Osterbrock2006}, which predicts H$\alpha$/H$\beta = 2.8$ and H$\alpha$/H$\gamma = 6.0$. Our analysis is not sensitive to the precise ratios here, and we obtain essentially the same results for any reasonable choices, though we get poor fits at early times if H$\alpha$/H$\beta \sim 1$. For He\,I, we use the model ratios from \citep{Benjamin1999}, which show little sensitivity to temperature or density: $\lambda5876 / \lambda4471 \approx 2.5$, $\lambda5876 / \lambda6678 \approx 5$.
This leaves a total of 9 free parameters. We fit to the spectra using the Optimize routine in \textsc{scipy}.

We show the fits in Figure \ref{fig:profiles} and plot the derived luminosities, velocity widths and shifts of H$\alpha$ and He\,II in Figure \ref{fig:linev} (He\,I contributes significantly only in the earliest spectrum).  He\,II exhibits a fairly flat luminosity with time, while H$\alpha$ fades by at least a factor 5. We therefore determine an initial ratio He\,II\,/\,H$\alpha\sim2$, but a markedly different ratio after 100 days of He\,II\,/\,H$\alpha>10$.
We measure an anomalously low He\,II luminosity in the 62 day spectrum, for which we cannot rule out an issue due to the low signal-to-noise ratio of the data.


At early times, lines are centred close to their rest wavelengths, but we find a substantial shift in the He\,II line at late times (after H$\alpha$ fades) measuring a maximum blueshift of $\sim 8000$\,\kms\ at 112 days. We find a smaller but still significant blueshift of $\sim 5000$\,\kms\ at 145 days, but these data are noisier. Such shifts are too large to be explained by the un-modelled Bowen lines (N\,III, C\,III), which are offset from He\,II only by 3000\,\kms. 

\begin{figure*}
\centering
	\includegraphics[width=12cm]{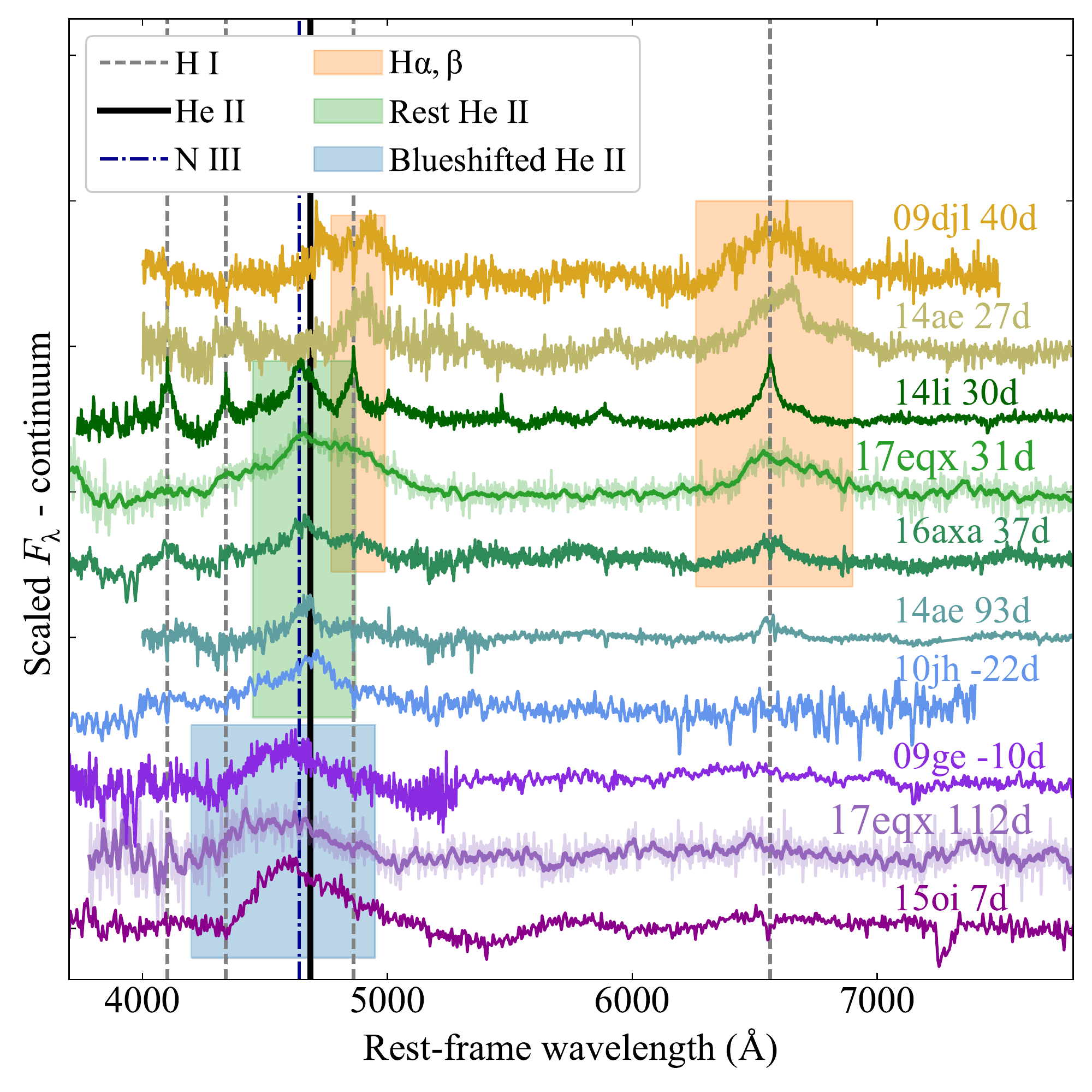}
    \caption{
        Spectroscopic comparison of AT2017eqx with other TDEs. All spectra have had continuum removed to highlight line emission. Phases are listed relative to maximum light or discovery. The early spectra of AT2017eqx show the same lines as the H/He-strong TDEs ASASSN-14li \citep{Holoien2016} and iPTF16axa \citep{Hung2017}, while late spectra match the He-strong TDEs with blueshifted lines PTF09ge \citep{Arcavi2014} and ASASSN-15oi \citep{Holoien2016b}. This demonstrates that a TDE can change its apparent spectral type, and that observed lines are more indicative of the physical conditions in the stellar debris rather than its composition. ASASSN-14ae \citep{Holoien2014} at first shows only hydrogen, but as these lines become weaker it also develops a He\,II line, suggesting a similar evolution to AT2017eqx.
        }
    \label{fig:speccomp}
\end{figure*}

\begin{figure*}
\centering
	\includegraphics[width=\columnwidth]{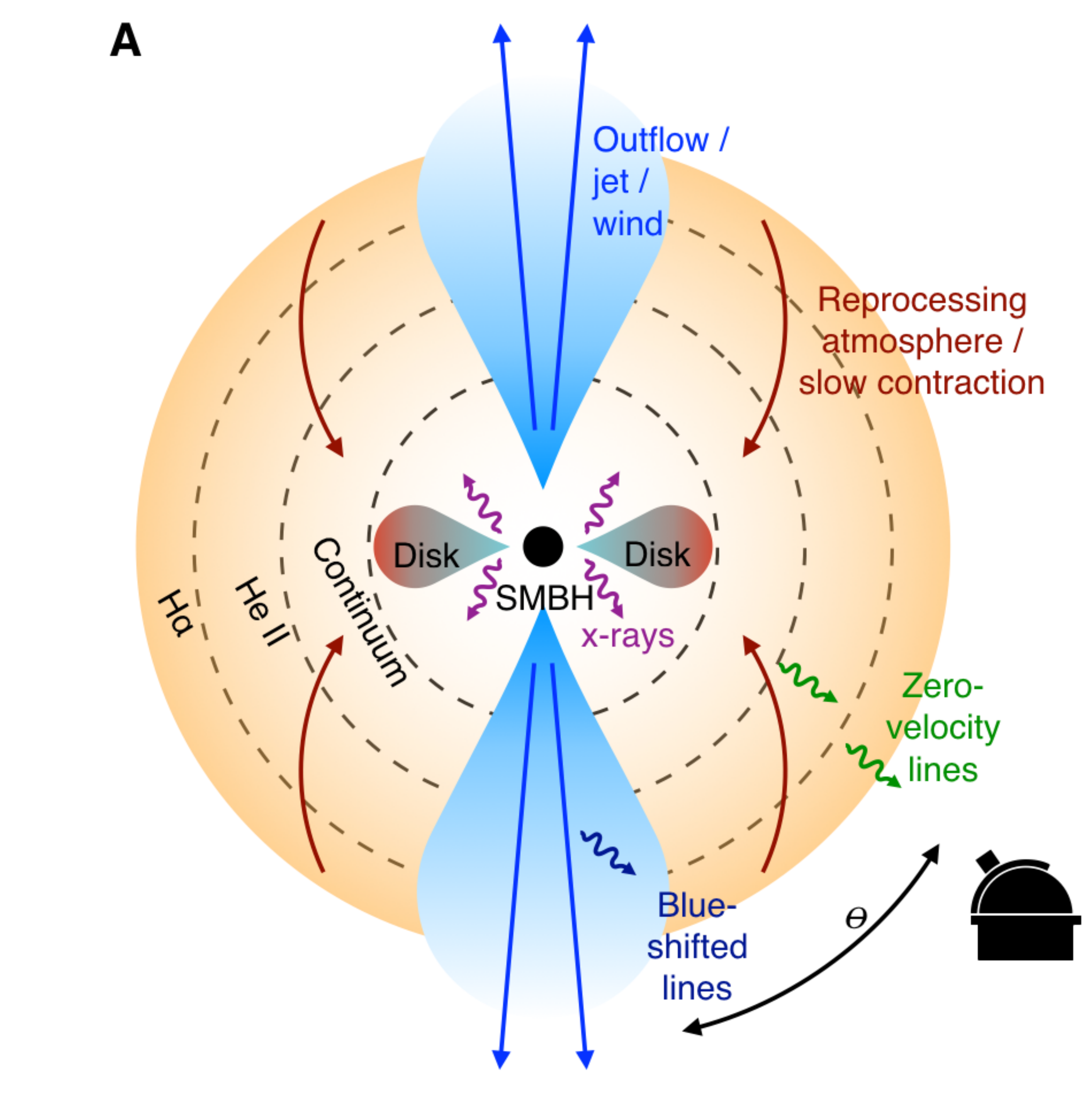}
	\includegraphics[width=\columnwidth]{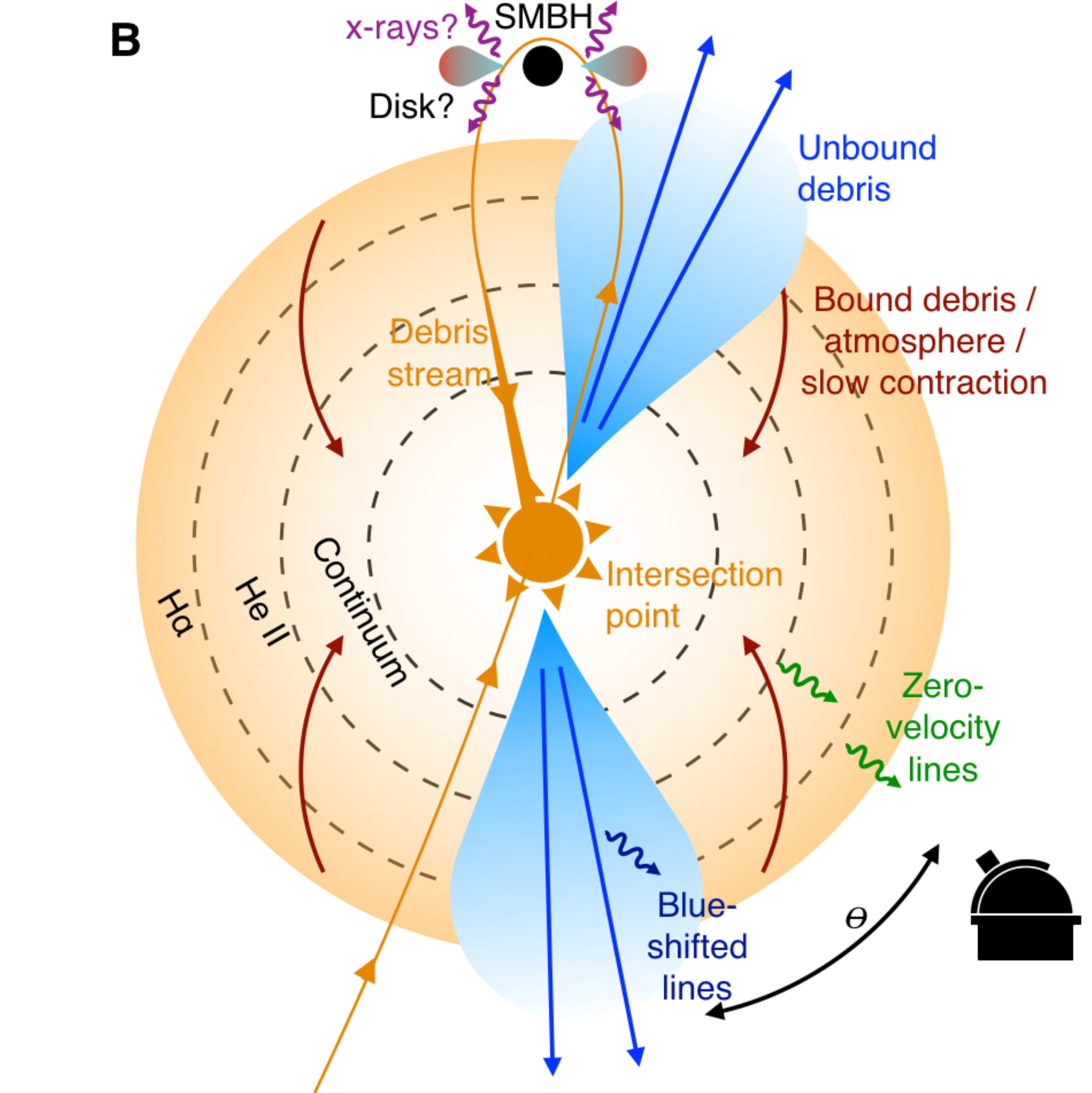}
    \caption{ Schematic of AT2017eqx and unified model for TDE line emission. Left (A): If an accretion disk forms quickly, X-rays heat a reprocessing layer powering the optical light. In the polar region, a wind can develop \citep{Dai2018,Metzger2016}. Right (B): Alternatively, energy is produced by collisions between debris streams. Bound debris forms an atmosphere \citep{Jiang2016}, while unbound material escapes along the stream directions, leading to a qualitatively similar inflow/outflow model \citep[see also][]{Hung2019}. In either scenario, extended atmospheres produce He\,II\,$\lambda4686$ and Balmer lines, whereas compact ones produce only He\,II \citep{Guillochon2014,Roth2016}. Outflows produce blueshifted emission lines \citep{Roth2018}, but are visible only for certain viewing angles. If the atmosphere contracts, blueshifted emission can be revealed to observers at larger angles. In case A, viewing angles far from the disk plane, which show blueshifted lines, may also reveal X-rays earlier, whereas no such correlation exists in case B.
}
    \label{fig:cartoon}
\end{figure*}

\subsection{Interpretation of the line evolution}\label{sec:interp}

The spectroscopic evolution of AT2017eqx described in the previous sections can be summarised in two distinct phases: at early times, it is strong in H\,I and He\,II, with lines centred close to zero velocity. At later times, beyond 60-100 days, the spectrum is dominated by a single broad feature close to He\,II but with a blueshift of up to $\sim 8000$\,\kms. Both of these morphologies have been observed in previous TDEs, but AT2017eqx is unique among the TDE sample to date in showing both a strongly evolving He\,II/H$\alpha$ ratio and a late-onset blueshift.

Given the surprising spectroscopic evolution in AT2017eqx, we compare to literature TDEs in Figure \ref{fig:speccomp}. We include H-strong TDEs PTF09djl \citep{Arcavi2014} and ASASSN-14ae \citep{Holoien2014}, the H/He-strong ASASSN-14li \citep{Holoien2016} and iPTF16axa \citep{Hung2017}, and the He-strong PS1-10jh \citep{Gezari2012}, PTF09ge \citep{Arcavi2014} and ASASSN-15oi \citep{Holoien2016b}; the latter latter two events show a significant blueshift in their He\,II line profiles. While AT2017eqx falls neatly into the H/He-strong group around the time of maximum, the late-time spectrum is a close match to the He-strong events, including the blueshift in He\,II.

The transition in AT2017eqx from a H-strong to H-poor spectrum confirms theoretical arguments that the difference between H-strong and He-strong TDE spectra is due to physical conditions rather than the composition of the disrupted star \citep{Guillochon2014,Roth2016}. The fact that AT2017eqx changes its spectral morphology on a relatively short timescale indicates that line formation is likely very sensitive to the precise configuration of the system.

Early models of TDE spectra suggested that large He\,II/H$\alpha$ ratios could be explained as a consequence of near-complete hydrogen ionization throughout the stellar debris \citep{Guillochon2014}. Such a situation implies that as the ionizing flux from the inner disk fades, hydrogen lines (and He\,I 5876) could become apparent \emph{later} in the TDE evolution -- the opposite of what we observe in AT2017eqx.
However, \citet{Roth2016} argued that under more realistic TDE conditions, wavelength-dependent optical depth is the most important factor in determining line strengths. Their radiative transfer calculations showed that -- all else being equal -- a more compact envelope gives a larger He\,II/H$\alpha$ ratio, because H$\alpha$ is self-absorbed at most radii whereas He\,II is thermalised at greater depth and therefore emitted over a greater volume. Thus the transition from a H-strong to He-strong spectrum can potentially be explained if there is a contraction of the envelope towards the SMBH. 

A contracting envelope is consistent with the observed decrease in luminosity at constant temperature\footnote{A reduction in photospheric radius is also possible even in outflowing material, if the density and ionization are decreasing as in a supernova, but this is generally accompanied by a decreasing temperature. Moreover, a lower density should favour H$\alpha$ production, as photons emitted from greater depths could escape without destruction by self-absorption.} (Figure \ref{fig:bol}). Moreover, our results from modelling the photometry with blackbody fits and \textsc{mosfit} indicated a photosphere that grows and shrinks in direct proportion to the luminosity. In the time between maximum light and the disappearance of the hydrogen lines, the luminosity (and therefore radius) decrease by roughly an order of magnitude, which could account for the change in the line ratios \citep{Roth2016}.

Equally important is the blueshift in He\,II. Blending with other lines such as He\,I or the Bowen lines may make a contribution, but these lines are not sufficiently blue to account for the size of the shift. One species that does emit at approximately the right wavelength is Fe\,II. \citet{Wevers2019b} identified these lines in AT2018fyk, and argued that they could also account for the apparently blueshifted He\,II line profile in TDEs like ASASSN-15oi. While we cannot rule out an Fe\,II contribution in AT2017eqx, these lines are thought to originate from dense gas close to a newly-formed accretion disk, and so would seem to be inconsistent with the lack of X-ray emission or other disk signatures in AT2017eqx.
If a disk was visible, another way to induce a blueshift is Doppler boosting of the blue (approaching) side. We disfavour this for two reasons: first, if the optical depth is low enough to reveal the disk, we would expect to see H$\alpha$, as in other events with disk-like line profiles \citep{Arcavi2014,Holoien2018}. Second, the disk should be hotter than the envelope, but we see no clear increase in temperature, nor the onset of X-ray emission.

\citet{Roth2018} explained the blueshifted profiles in TDEs as evidence for electron-scattered line emission in an outflowing gas. Yet we have previously shown that the increasing He\,II/H$\alpha$ ratio indicated a net inflow of material. Therefore to account for the full spectroscopic evolution of AT2017eqx requires both inflowing and outflowing gas, along with an appropriate geometry. We show a schematic of such a model in Figure \ref{fig:cartoon}.

In this scenario, the luminosity is generated in a small region, either from an accretion disk or at the intersection point between colliding debris streams. Initially, the observer sees (reprocessed) emission from a quasi-static atmosphere of bound debris \citep{Jiang2016}. Its extent, $\sim10^{14}$--$10^{15}$\,cm, is roughly proportional to the luminosity from the TDE engine. Following \citet{Roth2016}, the dominant emission lines depend on the extent and optical depth of this layer. Lines from this region are broadened by electron scattering, but centered at their rest-frame wavelength.

Above and below the plane of the disk, or parallel to the streams, an outflow forms -- either from a disk wind \citep{Metzger2016}, or material on unbound trajectories \citep{Jiang2016}. This produces emission lines with a net velocity shift \citep{Roth2018}. The crucial point is that whether the observer sees outflowing gas depends on whether their line of sight is obstructed by the envelope. But even for obstructed sight lines, outflows can eventually be revealed as they expand or the envelope contracts. Applying this to AT2017eqx, as the atmosphere shrinks and suppresses the H\,I emission, we are exposed to more of the outflowing material, causing the blueshift of He\,II. Thus this model naturally accounts for why an evolving He\,II/H$\alpha$ ratio is associated with a late-onset blueshift.

\subsection{Implications}

Our model for AT2017eqx suggests that for TDEs more generally, viewing angle may be the primary determinant as to whether we see inflowing or outflowing gas, the radial extent of which is important in setting the line ratios. Interpreting the spectroscopic diversity of TDEs as a manifestation of viewing similar sources from different angles evokes the unified model for active galactic nuclei \citep{Antonucci1993}, in which the diverse observational properties of AGN depend on our sight-line towards any associated jet, torus or broad/narrow-line regions. The term `unified model' has recently been applied to TDEs by \citet{Dai2018}, who showed using simulations how the diversity of observed TDE X-ray/optical ratios, temperatures and jets may vary as a function of viewing angle. While our interpretation of AT2017eqx and other TDEs has been developed independently, and applies primarily to the spectral lines, here we discuss how our model complements that work towards a unified model of TDEs.

In the accretion-powered paradigm, \citet{Dai2018} find that X-rays only escape for viewing angles close to the pole. In our model, such sight-lines are associated with blueshifted emission in the early spectra (or possibly disk profiles for very small angles). If most TDEs are powered by accretion, there should exist a correlation between blueshifted line profiles and detectable X-ray emission. 
In the case of stream-stream collisions, material may still ultimately accrete onto the SMBH and produce X-rays, but because the optical flare occurs off-center, we would expect no correlation between X-rays and blueshifted lines. Correlating the X-ray and optical properties of TDEs can therefore be used to test our proposed model, and possibly provide a way to determine whether most TDEs are powered by accretion or stream-stream collisions \citep[see also][]{Pasham2017}.

There are several complications to this picture, such as the non-spherical geometry of real TDEs, and the fact that powerful outflows can also produce X-rays (at least in some relativistic TDEs). Moreover, some TDEs show two light curve maxima, or a far-UV excess several years after disruption, which may indicate both accretion and stream collisions are at work \citep{Leloudas2016,Wevers2019b,vanVelzen2018c}. Searching for this correlation will therefore require large statistical samples of TDEs, but these will be provided soon by current and next-generation surveys.

\subsection{Application to other TDEs}

The schematic shown in Figure \ref{fig:cartoon} implies three possible scenarios for a typical TDE spectrum:
\begin{enumerate}
    \item The TDE is viewed approximately parallel to the disk (A) or orbital plane (B). An observer sees only reprocessed optical radiation from the envelope, with lines centered close to zero velocity. The ratio of He\,II/H$\alpha$ depends on the optical depth in this material, which may evolve as the debris expands or contracts in response to heating from below. 
    \item The TDE is viewed perpendicular to the disk (A) or orbital plane (B). This observer can see outflowing material, and spectral lines will have a net blueshift. If the power source is accretion onto the SMBH, and the outflow carves a cavity in the envelope, this could also be observed as an X-ray TDE.
    \item The TDE is viewed at an intermediate angle. This observer will most likely see only the envelope initially, but if this layer contracts, the outflow on the near side can be revealed leading to blue-shifted emission lines. X-rays may not become visible until much later. This scenario may explain the spectroscopic evolution seen in AT2017eqx.
\end{enumerate}
Here we consider some other TDEs within this context. Spectra have been obtained from the Open TDE Catalog \citep{Guillochon2017} and Weizmann Interactive Supernova Data Repository (WISeREP) \citep{Yaron2012}.

PS1-10jh, one of the earliest optical TDEs, showed only He\,II lines, but in this case without a blueshift even at 250 days. No X-rays were detected despite deep observations. Our model naturally accounts for this, as even if an accretion disk did form promptly, it may never become visible due to an approximately side-on viewing angle through the atmosphere ($\theta \sim 90^\circ$ in Figure \ref{fig:cartoon}), in agreement with other studies \citep{Dai2018}.

ASASSN-14ae \citep{Holoien2014,Brown2016} initially shows only hydrogen lines in its spectrum, but it gradually develops a He\,II line over time, while H$\alpha$ becomes weaker. By 93 days, He\,II is stronger than H$\alpha$. Thus ASASSN-14ae evolves from a H-strong to a mixed H/He spectrum, as shown in Figure \ref{fig:speccomp}, analogous to how AT2017eqx evolves from a H/He to a He-strong spectrum. Therefore this is another event consistent with a contracting envelope. ASASSN-14ae did not exhibit X-ray emission.

iPTF16fnl showed H\,I and He\,II in its early spectra \citep{Blagorodnova2017,Brown2018}, centered close to their rest-frame wavelengths, and the X-ray/optical ratio was constrained to be $<10^{-2}$ \citep[consistent with background fluctuations;][]{Auchettl2017,Blagorodnova2017}. Along with these similarities to AT2017eqx, this event may also show a comparable evolution in the He\,II\,/\,H$\alpha$ ratio, indicating the atmosphere could be contracting in a similar manner. However, no blueshifts are observed, suggesting a viewing angle closer to side-on.

ASASSN-15oi showed blueshifted He\,II around maximum light (Figure \ref{fig:speccomp}), and X-ray emission that gradually increased in time as the optical light faded \citep{Holoien2016b,Gezari2017,Holoien2018b}. In our model, seeing the blueshifted emission at early times implies a viewing angle close to the direction of the outflow, so if a disk had formed promptly we may have expected that this event would be X-ray bright from the beginning, rather than slowly increasing over a year. Thus if our proposed geometry applies to ASASSN-15oi, this could support the interpretation \citet{Gezari2017}: that the X-rays were a sign of gradual disk formation, and that the earlier optical emission was therefore collisional.

ASASSN-14li is one of the best-studied H- and He-strong TDEs \citep{Holoien2016,Brown2017}. The optical lines in this event are consistent with their rest-frame wavelengths (though He\,II may show a slight offset due to blending with N III; \citealt{Leloudas2019}). 
This event was also detected in X-rays and radio, with X-ray absorption lines and radio luminosity corresponding to a velocity of up to $\sim0.1c$ \citep{vanVelzen2016,Alexander2016,Kara2018}. However, other X-ray and UV lines are seen at much lower velocities, of a few hundred \kms\ \citep{Miller2015,Cenko2016}, implying material with a range of velocities. Based on cross-correlating the temporal evolution of the emission in the X-ray, optical and radio regimes, \citet{Pasham2017} and \citet{Pasham2018} argue that the X-rays are produced by an accretion disk that modulates the radio jet, while the optical emission comes from a stream intersection region further away. In the context of our model, seeing early-time X-ray emission with only low-velocity spectral lines would appear to be possible only in case B, supporting the collsional picture preferred by \citet{Pasham2018}.

Overall, it appears that the limited sample of optical TDEs can be accommodated within a picture like that presented in Figure \ref{fig:cartoon}: PS1-10jh, ASASSN-14ae, and iPTF16fnl are consistent with either accretion or stream collisions, but ASASSN-15oi and ASASSN-14li seem to favour the latter. However, to consistently explain the X-ray and optical properties together likely requires a combination of both processes on different timescales. This may make it difficult to identify the suggested correlation between X-ray emission and blueshifts until much larger TDE samples are available.

\section{Host galaxy}
\label{sec:host}

A surprisingly large fraction of TDEs have been found in a specific class of quiescent Balmer-strong absorption galaxies \citep{Arcavi2014,French2016,Graur2018,LawSmith2017}, defined spectroscopically by the presence of Balmer lines in absorption and a lack of nebular emission lines. This combination signifies that star formation was likely significant up to $\sim1$\,Gyr ago but has now largely ceased, leaving behind A type stars that dominate the stellar light (whereas the ionizing O and B stars have already died).

From our host galaxy spectrum, we measure a Lick index (the equivalent width of H$\delta$ using line and continuum bandpasses defined by \citealt{Worthey1997}) H$\delta_A = 1.85$\,\AA. We also find that H$\alpha$ is visible only in absorption, with an equivalent width of 3.1\,\AA. This combination satisfies the cut used by \citet{French2016}, encapsulating 75\% of TDE hosts known at the time, and only 2\% of SDSS galaxies. 
Using the largest TDE sample to date, a more recent study by \citet{Graur2018} found that the fraction of such host galaxies among optical TDEs is $\sim 33\%$, which is still a significant overabundance. The host of AT2017eqx adds to this over-representation of Balmer-strong absorption galaxies in the TDE host population.

The galaxy is also detected in a number of sky surveys with public catalogs. We retrieved $ugriz$ magnitudes from SDSS DR14 \citep{Abolfathi2018}, $grizy$ magnitudes from PanSTARRS DR1 \citep{Flewelling2016}, and the $W1$ magnitude from WISE \citep{Wright2010}, and fit the resultant host SED using \textsc{prospector} \citep{Leja2017} to derive physical parameters such as stellar mass and star-formation rate. The code includes the effects of stellar and nebular emission, metallicity, dust reprocessing, and a non-parametric star-formation history, within a nested sampling framework.

\begin{figure}
\centering
	\includegraphics[width=\columnwidth]{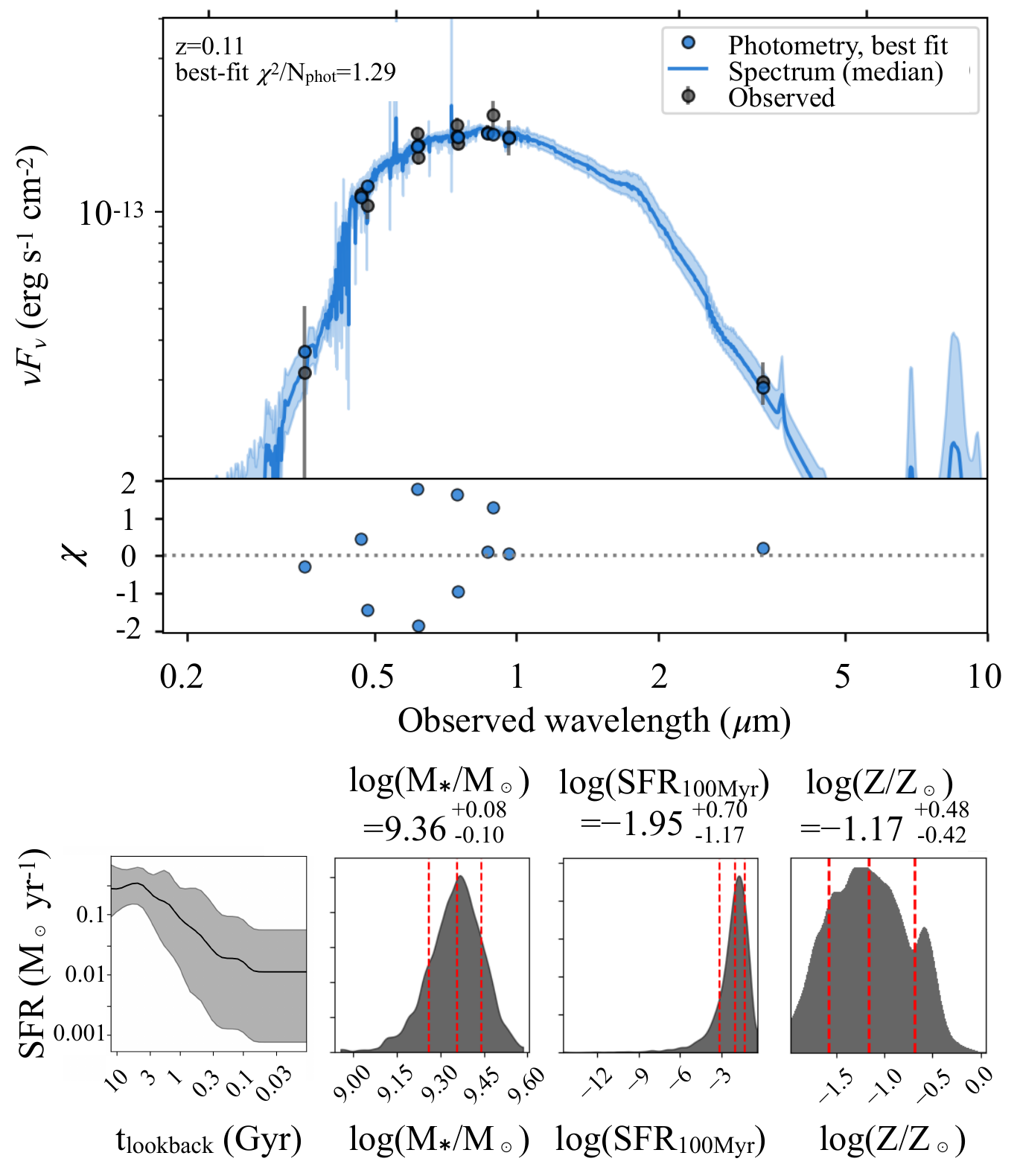}
    \caption{Fit to the archival SED of the AT2017eqx host galaxy using \textsc{prospector} \citep{Leja2017}. The lower panels show the star-formation history, and posteriors for stellar mass, present-day SFR, and metallicity. The recent decline in SFR is consistent with other TDE hosts.}
    \label{fig:prospector}
\end{figure}

The galaxy SED and \textsc{prospector} fitting results are shown in Figure \ref{fig:prospector}. A small offset is visible between the SDSS and PS1 photometry, but the model well captures the SED shape. We find a stellar mass $\log (M_*/M_\odot)= 9.36 \pm0.1$, with low star-formation rate and metallicity. The star-formation history is of particular interest: this drops by an order of magnitude over the past Gyr, consistent with expectations for the galaxy spectral type. Using the scaling relation between bulge stellar mass and SMBH mass from \citet{Kormendy2013}, and assuming that $M_{*{\rm ,bulge}}=M_*$ for an elliptical galaxy, we estimate a black hole mass of $\log (M_{\rm BH}/M_\odot)\approx6.8$, consistent with the results of fitting the TDE light curve.

The SMBH mass we infer for AT2017eqx is typical for optical TDE host galaxies, $\sim10^6$\,\M, as measured from stellar velocity dispersions by \citet{Wevers2019}. The same study found that X-ray selected TDEs have a flatter distribution in $M_{\rm BH}$ and typically show a smaller emitting region, consistent with an accretion disk rather than an inflated atmosphere. This may indicate that the dominant emission source can vary depending on the SMBH mass.

The host of AT2017eqx has a lower stellar mass and star-formation rate (SFR) than any of the galaxies studied by \citet{LawSmith2017}, with the possible exception of the host of RBS\,1032 \citep{Maksym2014}, whose origin as a TDE has been questioned \citep{Ghosh2006}. Typical TDE hosts have a SFR that is $\sim0.5$\,dex below the star-forming main sequence for a given mass. The host of AT2017eqx is similar, though with a slightly larger offset of $\approx 1$\,dex. This is consistent with the positions of other quiescent, Balmer strong galaxies \citep{French2016}.

\begin{figure}
\centering
	\includegraphics[width=\columnwidth]{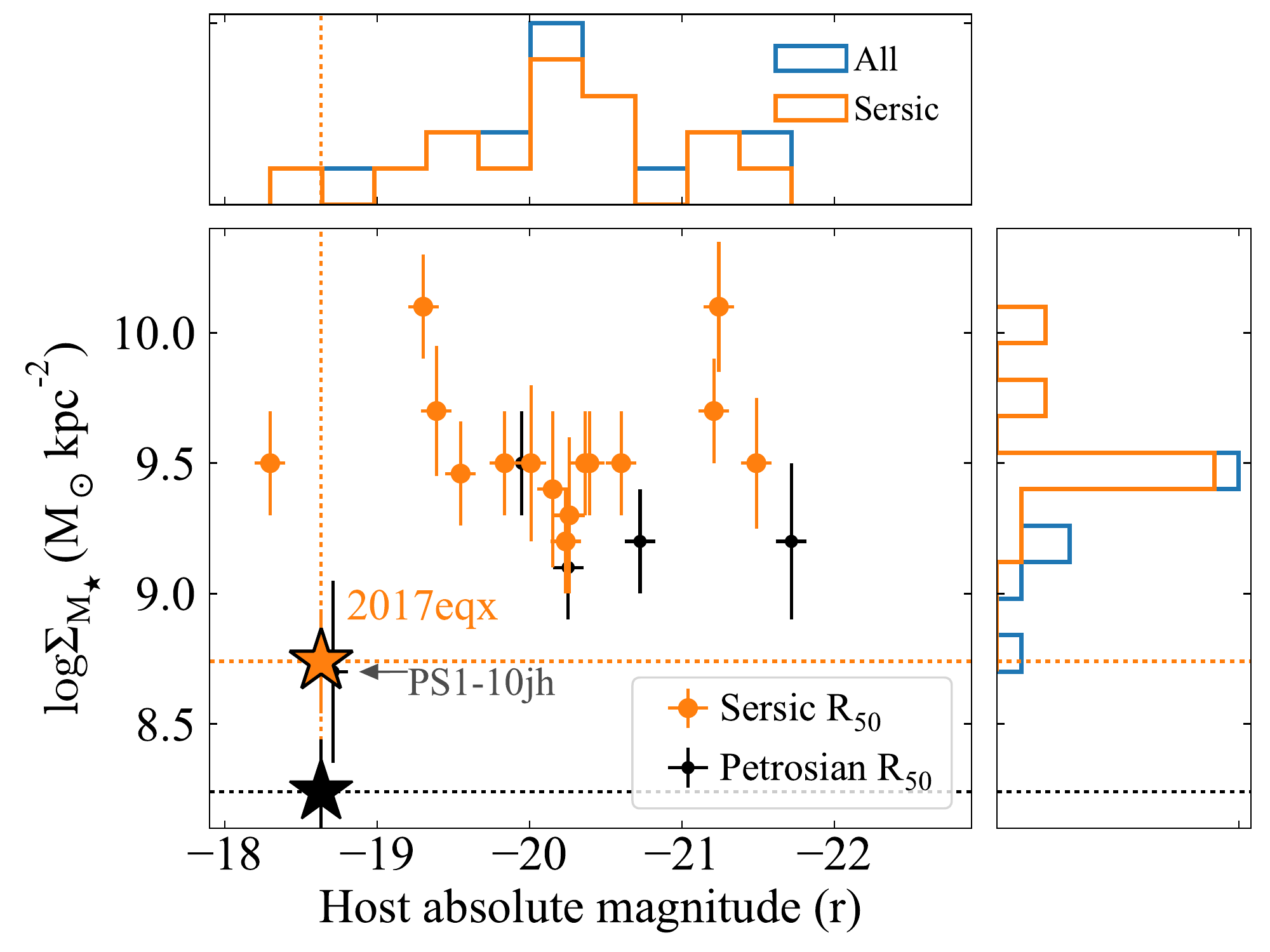}
	\includegraphics[width=\columnwidth]{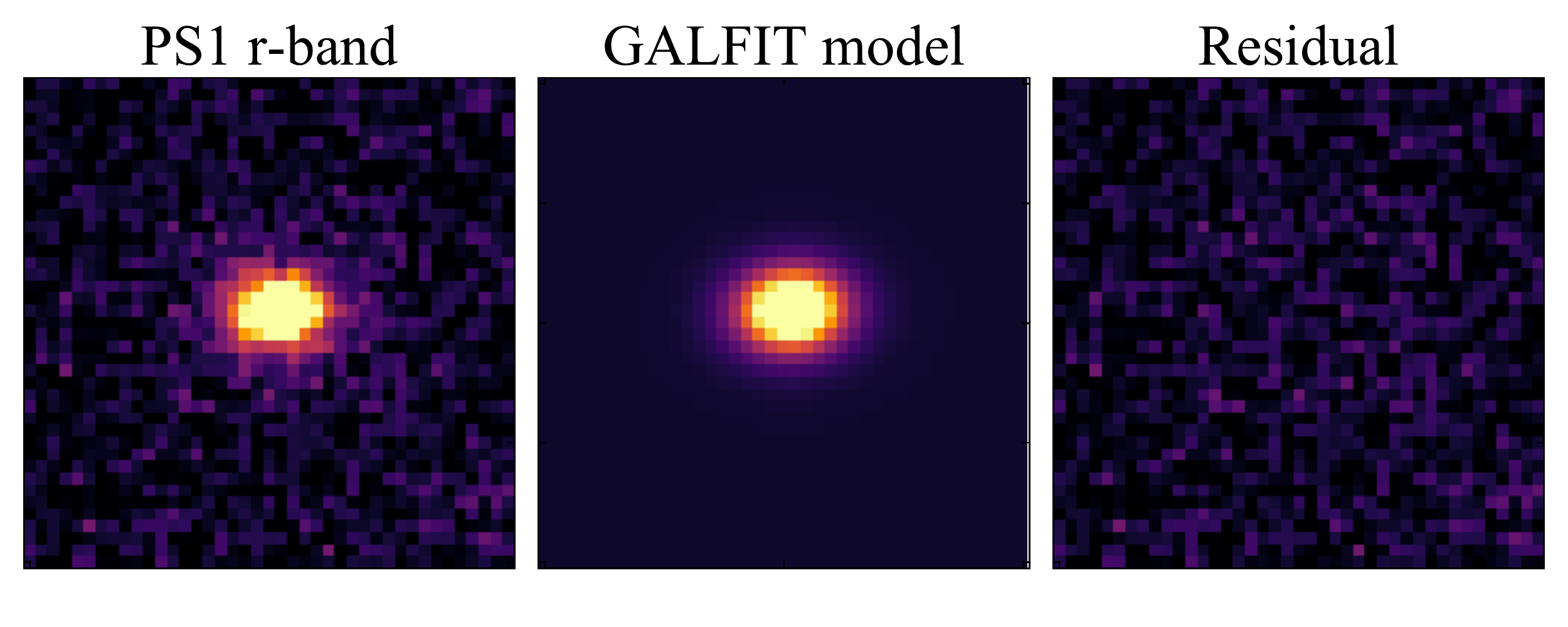}
    \caption{Stellar mass surface density within the half-light radius ($R_{50}$) versus $r$-band magnitude for the host of AT2017eqx and a TDE host comparison sample \citep{Graur2018}. AT2017eqx resides in a galaxy that is significantly less centrally-concentrated than typical TDE hosts, with only PS1-10jh falling in a similar part of the plot. The bottom panels show the GALFIT model used to determine the S\'ersic index and half-light radius.
    }
    \label{fig:sigma}
\end{figure}

TDE host galaxies also tend to have centrally-concentrated mass distributions \citep{Graur2018,LawSmith2017}, with an average S\'ersic index of $4.3^{+1.0}_{-1.9}$ , significantly higher than typical galaxies in the same mass range \citep{LawSmith2017}. We retrieved an $r$-band image of the host of AT2017eqx from PanSTARRS DR1, and fit the light profile using \textsc{galfit} \citep{Peng2002}. We find an excellent fit with a S\'ersic index of only 0.7 (Figure \ref{fig:sigma}), indicating a less sharply-peaked light distribution than typical TDE hosts.

\citet{Graur2018} showed that TDE hosts generally have a stellar mass surface density $\log(\Sigma_{M*}/(M_\odot\,{\rm kpc}^{-2}))>9$, which is consistent with typical quiescent galaxies, but high for star-forming TDE hosts. This can be interpreted as evidence for a high density of stars around the central SMBH, which could naturally lead to a higher rate of TDEs \citep{Graur2018}, as earlier proposed by \citep{Stone2016,Stone2016b}.
For consistency with that study, we recalculate the stellar mass of the AT2017eqx host galaxy by fitting only the SDSS magnitudes, using \textsc{lephare} \citep{Arnouts1999,Ilbert2009}. We find $\log M_*=9.19\pm0.17$, consistent with our results from \textsc{prospector}. To convert mass into surface density, we use the half-light radius from our S\'ersic fit: $1.82''=1.67$\,kpc, giving a surface density $\log(\Sigma_{M*})=8.74\pm0.2$. 

We plot this compared to other TDE hosts in Figure \ref{fig:sigma}. AT2017eqx resides in one of the faintest galaxies for known TDEs, and has an unusually low surface mass density. Only PS1-10jh has a host with comparable $\Sigma_{M*}$. However, this galaxy had a Petrosian half-light radius (from SDSS), rather than S\'ersic. Taking the Petrosian radius of the AT2017eqx host from SDSS, and applying the conversion $R_{50}=R_{\rm petro}-0.3''$ \citep{Graur2018}, we find an even lower stellar surface mass density of $\log(\Sigma_{M*})=8.24\pm0.2$, significantly offset from other TDE hosts. 

The low surface mass density compared to other TDE hosts is consistent with the lower S\'ersic index. The surface mass density is also low compared to average quiescent galaxies, but is typical for star-forming galaxies in the volume-weighted sample from \citet{Graur2018}. Taking their empirical relation between TDE rate and $\Sigma_{M*}$, we find that $\lesssim 10\%$ of TDEs are expected to occur in a galaxy with this surface density. We therefore conclude that although the host of AT2017eqx is somewhat unusual for TDEs, it is not overwhelmingly improbable to find a TDE in such a galaxy.

\section{Conclusions}
\label{sec:conc}

We have presented an in-depth study of a new TDE, AT2017eqx, discovered by the PanSTARRS Survey for Transients and ATLAS. We followed up this event with optical photometry and spectroscopy, UV photometry from \textit{Swift}, X-ray imaging and \textit{Chandra} and radio observations with the VLA. The SED maintains a roughly constant colour temperature of $\gtrsim 20,000$\,K for at least 150 days.

Non-detections in the radio indicate that AT2017eqx did not launch a relativistic jet, and constrain the luminosity of any slower outflow to at most the level seen in ASASSN-14li. Our X-ray limits are among the deepest for any TDE to date. If an accretion disk formed promptly after the disruption, the system would have to remain optically thick to X-rays for at least 500 days.

Modelling of the UV and optical light curve with \textsc{mosfit} suggests that AT2017eqx resulted from the complete disruption of a solar-mass star by a black hole of $\gtrsim 10^{6.3}$\,\M, with no significant viscous delay, similar to other UV-optical TDEs \citep{Mockler2018}. This SMBH mass is consistent with the observed properties of the host galaxy, with SED fitting indicating a host stellar mass $\approx 10^{9.3}$\,\M. The inferred star-formation history, and analysis of a galaxy spectrum, show that this is yet another Balmer-strong absorption galaxy hosting a TDE. However, it is one of the least massive TDE hosts to date, and the stellar mass surface density is relatively low compared to typical TDE hosts \citep{Graur2018}.

The most important new results come from the spectroscopic evolution of AT2017eqx. In the first months following the light curve maximum, the spectrum shows broad emission lines from He\,II $\lambda4686$ and the Balmer series, with some evidence for N\,III Bowen flourescence \citep{Blagorodnova2018}, and He\,I in the earliest spectrum. The H\,I and He\,II lines initially have widths of $\gtrsim 20,000$\,\kms, but are centered close to their rest-frame wavelengths. This resembles other TDEs such as ASASSN-14li, iPTF16axa and AT2018dyb. However, between 60-100 days after maximum, the H\,I lines disappear, while He\,II develops a blueshift of $\sim 5000$-8000\,\kms, resembling He-strong TDEs such as PTF09ge and ASASSN-15oi. Such a stark transition has not been seen in TDEs before, and it confirms theoretical arguments that a lack of H\,I lines in the spectrum of a TDE does not indicate a low abundance of hydrogen in the disrupted star.

We propose that the evolution can be explained if the debris around AT2017eqx consist of a quasi-spherical envelope, and a relatively narrow outflow from a disk wind or unbound debris. Viewing from an intermediate angle (i.e. neither parallel nor perpendicular to the outflow), we initially see emission lines of H\,I and He\,II from the envelope at zero velocity. As this layer contracts, either due to simple fallback or in response to decreasing radiation pressure, the increasing optical depth suppresses hydrogen emission. At the same time, we see more of the outflow, which produces the observed blueshift in the He\,II line at late times. 
The picture we have proposed for AT2017eqx is compatible with many TDEs from the current available samples, though these are limited in size. It has recently been shown that many TDEs exhibit outflow signatures in their UV spectra \citep{Hung2019}.
Future work should determine whether it can also be applied other TDEs with featureless spectra or absorption lines \citep{Cenko2012b,Chornock2014,Leloudas2016,Blanchard2017}, or with disk-like line profiles \citep{Holoien2018}. 

Furthermore, determining whether our model applies to both X-ray weak and strong TDEs will crucial to understanding TDE physics and geometry.
We suggest that searching for a correlation between X-rays and blueshifted spectral lines in TDEs is a promising avenue to distinguish between reprocessed accretion power and stream-stream collisions as the dominant power source. However, this picture quickly becomes more complicated if many collision-powered optical TDEs also form X-ray accretion disks on relatively short timescales.
Identifying such a correlation between X-ray and optical properties will therefore require a much larger statistical sample, but this can soon be provided by the many ongoing and planned sky surveys such as ASASSN, ATLAS, PanSTARRS, ZTF and LSST.

\section*{Acknowledgements}

We thank an anonymous referee for many helpful comments that improved this paper. Thanks to Yuri Beletsky for IMACS observing, Jabran Zahid for help with the galaxy analysis, Jorge Anais, Jaime Vargas, Abdo Campillay and Nahir Mu{\~n}oz Elgueta for Swope observing, and Dave Coulter for writing the Swope scheduling software. We also thank Joel Aycock, Percy Gomez, and Yen-Chen Pan for assistance with Keck/LRIS observations.
M.N.~is supported by a Royal Astronomical Society Research Fellowship. The Berger Time-Domain Group is supported in part by NSF grant AST-1714498 and NASA grant NNX15AE50G. We acknowledge Chandra Award 20700239. K.D.A.~acknowledges NASA Hubble Fellowship grant HST-HF2-51403.001. ATLAS acknowledges NASA grants NN12AR55G, 80NSSC18K0284, and 80NSSC18K1575. S.J.S.~acknowledges STFC Grants ST/P000312/1 and ST/N002520/1.  O.G.~and J.L.~are supported by NSF Astronomy and Astrophysics Fellowships under awards AST-1602595 and AST-1701487.
The UCSC team is supported in part by NASA grant NNG17PX03C, NSF grant AST-1518052, the Gordon \& Betty Moore Foundation, the Heising-Simons Foundation, and by a fellowship from the David and Lucile Packard Foundation to R.J.F.
Data were obtained via the Swift archive, the Smithsonian Astrophysical Observatory OIR Data Center, the MMT Observatory of the Smithsonian Institution and the University of Arizona, and Las Campanas Observatory. NRAO is a facility of the NSF operated by Associated Universities, Inc. ATLAS products are made possible by the University of Hawaii, Queen's University Belfast, the Space Telescope Science Institute, and the South African Astronomical Observatory. 
ome of the data presented herein were obtained at the W. M. Keck Observatory, which is operated as a scientific partnership among the California Institute of Technology, the University of California, and NASA; the observatory was made possible by the generous financial support of the W. M. Keck Foundation. The authors wish to recognize and acknowledge the very significant cultural role and reverence that the summit of Mauna Kea has always had within the indigenous Hawaiian community. We are most fortunate to have the opportunity to conduct observations from this mountain.

\bibliographystyle{mnras}
\bibliography{at2017eqx-bib}



\begin{table*}
    \centering
    \caption{Log of photometric observations}
    \begin{tabular}{cccccccccccc}
    MJD & Phase (d) & Magnitude & Error & Band & Telescope & MJD & Phase (d) & Magnitude & Error & Band & Telescope \\
    \hline
57905.6	&	-14.5	&	20.14	&	0.18	&	o	&	ATLAS	&	57978	&	50.9	&	20.03	&	0.34	&	U	&	Swift/UVOT	\\
57911.6	&	-9.1	&	19.7	&	0.04	&	i	&	PS1	&	57978	&	50.9	&	19.39	&	0.24	&	W1	&	Swift/UVOT	\\
57926.6	&	4.5	&	19.57	&	0.14	&	o	&	ATLAS	&	57978	&	50.9	&	19.28	&	0.18	&	M2	&	Swift/UVOT	\\
57932.6	&	9.9	&	19.99	&	0.12	&	o	&	ATLAS	&	57978	&	50.9	&	18.83	&	0.11	&	W2	&	Swift/UVOT	\\
57936.6	&	13.5	&	20.03	&	0.14	&	o	&	ATLAS	&	57979.2	&	52	&	21.1	&	0.13	&	g	&	Swope	\\
57938.5	&	15.3	&	20.09	&	0.03	&	i	&	PS1	&	57979.2	&	52	&	21.16	&	0.14	&	r	&	Swope	\\
57945	&	21.1	&	19.05	&	0.2	&	U	&	Swift/UVOT	&	57979.2	&	52	&	21.25	&	0.13	&	i	&	Swope	\\
57945	&	21.1	&	18.82	&	0.18	&	W1	&	Swift/UVOT	&	57981.4	&	53.9	&	20.71	&	0.21	&	B	&	Keck/LRIS	\\
57945	&	21.1	&	18.45	&	0.13	&	M2	&	Swift/UVOT	&	57981.5	&	53.9	&	20.76	&	0.21	&	V	&	Keck/LRIS	\\
57945	&	21.1	&	18.28	&	0.09	&	W2	&	Swift/UVOT	&	57981.4	&	53.9	&	20.82	&	0.19	&	R	&	Keck/LRIS	\\
57951.3	&	26.8	&	20.47	&	0.05	&	g	&	Magellan/IMACS	&	57981.5	&	53.9	&	21.01	&	0.18	&	I	&	Keck/LRIS	\\
57951.3	&	26.8	&	20.96	&	0.04	&	r	&	Magellan/IMACS	&	57997.4	&	68.4	&	21.54	&	0.13	&	g	&	FLWO48/Kepcam	\\
57951.3	&	26.8	&	20.93	&	0.05	&	i	&	Magellan/IMACS	&	57997.4	&	68.4	&	21.89	&	0.24	&	r	&	FLWO48/Kepcam	\\
57951.3	&	26.8	&	21.31	&	0.05	&	z	&	Magellan/IMACS	&	57997.4	&	68.4	&	21.31	&	0.2	&	i	&	FLWO48/Kepcam	\\
57952.6	&	27.9	&	20.49	&	0.23	&	o	&	ATLAS	&	58002	&	72.5	&	19.66	&	0.49	&	W1	&	Swift/UVOT	\\
57953	&	28.3	&	19.55	&	0.27	&	U	&	Swift/UVOT	&	58002	&	72.5	&	19.01	&	0.28	&	M2	&	Swift/UVOT	\\
57953	&	28.3	&	19.06	&	0.2	&	W1	&	Swift/UVOT	&	58002	&	72.5	&	19.49	&	0.28	&	W2	&	Swift/UVOT	\\
57953	&	28.3	&	18.75	&	0.15	&	M2	&	Swift/UVOT	&	58006	&	76.1	&	19.56	&	0.29	&	M2	&	Swift/UVOT	\\
57953	&	28.3	&	18.33	&	0.09	&	W2	&	Swift/UVOT	&	58006	&	76.1	&	19.32	&	0.22	&	W2	&	Swift/UVOT	\\
57956.3	&	31.3	&	20.59	&	0.04	&	g	&	Magellan/LDSS	&	58010.2	&	79.9	&	21.63	&	0.09	&	g	&	FLWO48/Kepcam	\\
57956.3	&	31.3	&	20.88	&	0.08	&	r	&	Magellan/LDSS	&	58012	&	81.5	&	19.62	&	0.24	&	M2	&	Swift/UVOT	\\
57956.3	&	31.3	&	21.12	&	0.11	&	i	&	Magellan/LDSS	&	58012	&	81.5	&	19.57	&	0.21	&	W2	&	Swift/UVOT	\\
57956.3	&	31.3	&	20.92	&	0.35	&	z	&	Magellan/LDSS	&	58012.2	&	81.7	&	21.63	&	0.26	&	g	&	Swope	\\
57956.5	&	31.5	&	20.21	&	0.21	&	o	&	ATLAS	&	58012.2	&	81.7	&	$>$21.00	&		&	r	&	Swope	\\
57964.5	&	38.7	&	20.7	&	0.22	&	o	&	ATLAS	&	58012.2	&	81.7	&	21.62	&	0.2	&	i	&	Swope	\\
57965.5	&	39.6	&	20.78	&	0.13	&	g	&	FLWO48/Kepcam	&	58019.1	&	87.9	&	21.85	&	0.32	&	g	&	Swope	\\
57965.5	&	39.6	&	20.8	&	0.11	&	g	&	FLWO48/Kepcam	&	58019.1	&	87.9	&	$>$21.26	&		&	r	&	Swope	\\
57965.5	&	39.6	&	21.08	&	0.14	&	r	&	FLWO48/Kepcam	&	58019.1	&	87.9	&	21.97	&	0.2	&	i	&	Swope	\\
57965.5	&	39.6	&	21.05	&	0.19	&	i	&	FLWO48/Kepcam	&	58028.1	&	96.1	&	21.82	&	0.15	&	g	&	FLWO48/Kepcam	\\
57968	&	41.8	&	20.05	&	0.37	&	U	&	Swift/UVOT	&	58028.1	&	96.1	&	22.08	&	0.14	&	r	&	FLWO48/Kepcam	\\
57968	&	41.8	&	19.13	&	0.21	&	W1	&	Swift/UVOT	&	58028.1	&	96.1	&	$>$22.01	&		&	g	&	Swope	\\
57968	&	41.8	&	18.99	&	0.17	&	M2	&	Swift/UVOT	&	58028.1	&	96.1	&	$>$21.50	&		&	r	&	Swope	\\
57968	&	41.8	&	18.64	&	0.1	&	W2	&	Swift/UVOT	&	58028.1	&	96.1	&	22.25	&	0.24	&	i	&	Swope	\\
57973	&	46.4	&	19.22	&	0.45	&	W1	&	Swift/UVOT	&	58035.1	&	102.4	&	$>$22.25	&		&	g	&	Swope	\\
57973	&	46.4	&	18.89	&	0.28	&	M2	&	Swift/UVOT	&	58035.1	&	102.4	&	$>$21.56	&		&	r	&	Swope	\\
57973	&	46.4	&	19.07	&	0.26	&	W2	&	Swift/UVOT	&	58035.1	&	102.4	&	$>$22.36	&		&	i	&	Swope	\\
    \hline
\end{tabular}
    Note: ATLAS photometry has been shifted by $+0.31$\,mag to match the $o-i$ colours inferred from spectroscopy
    \label{tab:phot}
\end{table*}

\begin{table*}
    \centering
    \caption{Log of spectroscopic observations}
    \begin{tabular}{ccccccc}
    MJD & Phase (d) & Telescope & Instrument & Grating & Exposure (s) & Airmass \\
    \hline
    57933.4 & 10  & MMT & Blue Channel & 300GPM & 900 & 1.1 \\
    57953.2 & 28  & Magellan & IMACS & G300-17.5 & 1500 & 2.1 \\
    57956.3 & 31  & Magellan & LDSS3c & VPH-all & 3$\times$1200 & 1.5 \\
    57991.3 & 62  & MMT & Blue Channel & 300GPM & 1800 & 1.1 \\
    58046.0 & 112 & Magellan & IMACS & G300-17.5 & 4$\times$1200 & 1.5 \\
    58083.1 & 145 & MMT & Blue Channel & 300GPM & 3$\times$1200 & 1.1 \\
    58277.4 & 321 & MMT & Binospec & 270 & 1800 & 1.3 \\
    \hline
    \end{tabular}
    \label{tab:spec}
\end{table*}

\begin{table*}
    \centering
    \caption{Log of VLA radio observations. Upper limits correspond to 3$\sigma$.}
    \begin{tabular}{ccccc}
    MJD & Phase (d) & Frequency (GHz) & Image RMS ($\mu$Jy) & Flux ($\mu$Jy) \\
    \hline
    57948.5  & 25  & 6.0  & 9.0 & $<27$ \\
    57948.5  & 25  & 21.7 & 27.0 & $<80$ \\
    57987.5  & 61  & 6.0  & 8.8 & $<26$ \\
    57987.5  & 61  & 21.7 & 25.0 & $<76$ \\
    Combined & 43  & 6.0 & 6.0 & $<18$ \\
    Combined & 43  & 21.7 & 17.0 & $<50$ \\
    \hline
    \end{tabular}
    \label{tab:radio}
\end{table*}

\begin{table*}
    \centering
    \caption{Log of X-ray observations. Upper limits correspond to 3$\sigma$.}
    \begin{tabular}{ccccccc}
    MJD & Phase & Telescope & Instrument & Duration & Count rate & 0.3-10\,keV flux \\
     & (d) &  &  & (ks) & (s$^{-1}$) & (\ergs\,cm$^-2$) \\
    \hline
    57162.5 &  & XMM & MOS-1 & 14.1 & $<1.95\times10^{−2}$ & $<1.9\times10^{-14}$ \\
    \hline
    57951.5 & 27  & Swift & XRT & 5.7 & $<1.91\times10^{−3}$ & $<7.9\times10^{-14}$ \\
    57981.5 & 54  & Chandra & ACIS-S & 10 & $<2.01\times10^{−4}$ & $<3.1\times10^{-15}$ \\
    58487.5 & 510  & Chandra & ACIS-S & 10 & $<2.22\times10^{−4}$ & $<2.6\times10^{-15}$ \\
    \hline
    \end{tabular}
    \label{tab:x}
\end{table*}


\bsp	
\label{lastpage}
\end{document}